\pdfoutput=1
\documentclass[sigconf,nonacm]{acmart}

\renewcommand\footnotetextcopyrightpermission[1]{}
\usepackage{booktabs}
\usepackage{multirow}
\usepackage{xcolor}
\usepackage{subcaption}
\usepackage{siunitx}
\usepackage{amsmath}
\usepackage{bookmark}
\usepackage{hyperref}

\newcommand{\perfboost}{\texttt{\mbox{CUDA}\_\hspace{0pt}\mbox{DISABLE}\_\hspace{0pt}\mbox{PERF}\_\hspace{0pt}\mbox{BOOST}}}

\begin{document}

\title{The Model Parking Tax: Configuration-Induced Idle Power in GPU LLM Inference Serving}

\author{Sai Sathvik Vadari}
\orcid{0009-0000-5952-7868}
\affiliation{
    \institution{\href{https://www.8bit.ai/}{8bit.ai}}
    \city{Hyderabad}
    \country{India}
}

\begin{abstract}
LLM inference services keep models resident in GPU memory around the clock to avoid cold starts, and treat the resulting idle power as a fixed cost of readiness. Much of it is not fixed. Combining 18~days of production telemetry (\num{335267} idle samples, 14~H100 GPUs) with controlled experiments on four architectures (H100, A100, L40S, B200), we show that idle power depends not on the resident model but on whether a CUDA context exists: the context forces a DVFS transition that holds the SM clock at full speed at 0\% utilization, adding 26--67\,W per GPU, while allocated VRAM has no measurable effect ($|\beta| < 0.02$\,W/GB). A single driver flag, \perfboost, recovers roughly 95--100\% of this overhead, with no measured latency penalty on H100 and A100 and no throughput penalty in an H100 load test. The overhead scales with the number of contexts: on a 4$\times$H100 node it adds ${\sim}52$\,W per GPU under tensor, data and mixed parallelism. At data-center scale, a mid-range scenario puts the recoverable energy near \num{460}\,GWh per year (${\sim}180$\,kt CO$_2$).
\end{abstract}


\maketitle

\section{Introduction}

The default operating model for AI inference is to deploy models to GPUs and keep them loaded in VRAM indefinitely, accepting continuous idle power consumption to avoid cold-start latency. Both sides of this tradeoff are individually well-studied. The cold-start problem has driven systems like ServerlessLLM~\cite{serverlessllm} (3.6--8.2$\times$ faster loading), NVIDIA Run:ai Model Streamer~\cite{nvidia-streamer} ($\sim$5\,s for Llama~3 8B), and checkpoint-optimized loaders~\cite{hydraserve,lambdascale}. The other side---the ongoing energy cost of readiness---has been treated as an unquantified constant.

The 2024 US Data Center Energy Usage Report estimates GPU idle power at $\sim$20\% of Thermal Design Power (TDP)~\cite{doe-report}. Patel et al.~\cite{patel-cal23} showed that CUDA contexts on A100 GPUs force higher clock frequencies even when idle. Concurrent work by Lei et al.~\cite{lei-idle} characterizes the prevalence of ``execution-idle'' at cluster scale---19.7\% of in-execution time and 10.7\% of energy across 756~GPUs over 31~days---but does not decompose idle power into its component causes. No prior study has measured the marginal energy cost \emph{per GB of VRAM allocation}---the quantity needed for rational keep-warm decisions---nor tested whether the result generalizes across GPU architectures and memory technologies. Moreover, modern LLM serving routinely distributes models across multiple GPUs via tensor parallelism (TP), data parallelism (DP), or both. If the parking tax multiplies linearly with~$N$, a single 4-GPU node wastes ${\sim}215$\,W idle; an 8-GPU node wastes ${\sim}430$\,W. At cluster scale with thousands of GPUs, even modest idle fractions translate to kilowatts of waste. Yet no study has measured whether the tax compounds, remains constant, or interacts with inter-GPU communication (NCCL).

This paper addresses that gap. We ask whether storing model weights in GPU VRAM incurs a measurable marginal idle energy cost, and whether the answer depends on GPU architecture. We answer with four complementary studies:

\begin{enumerate}
    \item \textbf{Production telemetry} (Phase~1): \num{335267} idle-state samples from 14~H100 GPUs across 2~nodes over 18~days, with VRAM allocations spanning 3\,MB to 77\,GB.
    \item \textbf{Controlled experiments} (Phase~2): Within-subject dose-response studies on four GPU architectures---H100 (HBM3), A100 (HBM2e), L40S (GDDR6), and B200 (Blackwell, HBM3e)---systematically varying VRAM from 0 to 40--72\,GB, validated with real model weights.
    \item \textbf{Vendor mitigation} (Phase~3): Evaluation of NVIDIA's \perfboost{} driver flag on H100 (Hopper) and A100 (Ampere), measuring idle power elimination, steady-state latency, and first-request DVFS ramp under both synthetic and vLLM workloads.
    \item \textbf{Multi-GPU scaling} (Phase~4): 9~conditions on a 4$\times$H100 SXM node testing TP=2, TP=4, DP=2, and TP=2$\times$DP=2 with and without \perfboost, measuring idle power, warm latency, DVFS decay, and cold-start ramp per parallelism strategy.
\end{enumerate}

Our central finding is that idle GPU power is \emph{piecewise constant} across all four architectures. We model idle power as:
\begin{equation}
    P_{\text{idle}}(C, V) = P_{\text{base}} + \Delta P_{\text{DVFS}} \cdot \mathbf{1}_{[C=1]} + \beta V
    \label{eq:model}
\end{equation}
where $P_{\text{base}}$ is the bare-idle power (no CUDA context), $\Delta P_{\text{DVFS}}$ is the discrete power step from the DVFS transition, $C \in \{0,1\}$ indicates a CUDA context, $V$ is VRAM allocated (GB), and $\beta$ is the marginal VRAM power coefficient. On all four devices, $|\beta| < 0.02$\,W/GB and we cannot reject $\beta = 0$ as a physically meaningful effect. The optimization priority is:
\begin{equation}
    \frac{\partial P}{\partial C} \gg \frac{\partial P}{\partial V}
    \label{eq:priority}
\end{equation}
This holds across HBM3, HBM2e, GDDR6, and HBM3e in our experiments---consistent across all tested architectures rather than an artifact of any single device.

\section{Background}

\subsection{GPU Power Architecture and DVFS}

NVIDIA datacenter GPUs support multiple power states via Dynamic Voltage and Frequency Scaling (DVFS). Without a CUDA context, streaming multiprocessor (SM) clocks drop to low-power idle (210--345\,MHz depending on architecture). Once a CUDA context is created---by any CUDA runtime call---the SM clock jumps to maximum boost and \emph{remains there} at 0\% utilization~\cite{patel-cal23}. This inverts the behavior one expects from a DVFS governor:
\begin{equation}
    \text{Expected: } P \downarrow \text{ as } U \to 0 \qquad \text{Observed: } P = \text{const when } C = 1
    \label{eq:dvfs-anomaly}
\end{equation}
where $U$ is SM utilization and $C=1$ denotes a resident CUDA context. This converts a \emph{variable} cost (expected) into a \emph{fixed} cost (observed), decoupling idle power from both utilization and memory occupancy. The behavior is a deliberate latency-oriented default rather than a defect (\S\ref{sec:default-regime}); we refer to it neutrally as \emph{boost-on-context}.

\subsection{GPU Memory Power Model}

GPU memory power decomposes into three components~\cite{hbm-spec,micron-ddr}:
\begin{equation}
    P_{\text{mem}} = P_{\text{background}} + P_{\text{refresh}} + P_{\text{activate}}
    \label{eq:mem-power}
\end{equation}
$P_{\text{background}}$ is constant whenever the memory clock runs. $P_{\text{refresh}}$ is proportional to \emph{capacity}, not occupancy---refresh circuitry sweeps all rows regardless of data content. $P_{\text{activate}}$ (row open/close) is zero at idle. This decomposition applies to both High Bandwidth Memory (HBM; HBM3, HBM2e, HBM3e) and GDDR6, as all are Dynamic Random-Access Memory (DRAM)-based with mandatory refresh. Therefore, at idle:
\begin{equation}
    \frac{\partial P_{\text{mem}}}{\partial \text{occupancy}} \approx 0
    \label{eq:mem-prediction}
\end{equation}
Memory controller and background power dominate at idle, remaining constant regardless of occupancy. DRAM refresh sweeps all rows per JEDEC specification, independent of which rows contain valid data---making $\beta \approx 0$ a hardware-physics prediction, not merely an empirical observation. A flat VRAM dose-response is therefore predicted by memory physics, and the invariance should be expected to generalize across DRAM technologies rather than being specific to one device. Our experiments test this prediction across four memory technologies.

\subsection{Related Work}

Patel et al.~\cite{patel-cal23} studied idle-but-allocated A100 and RTX~6000 Ada GPUs, finding DVFS locks at maximum frequency ($\sim$20\,W above bare idle). They did not vary VRAM levels, test cross-architecture invariance, or apply formal equivalence testing to bound the VRAM effect. We extend their result in three directions: VRAM independence as a first-class result with Two One-Sided Tests (TOST) equivalence bounds, cross-memory-technology invariance (HBM3, HBM2e, GDDR6, HBM3e), and evaluation of the vendor-provided \perfboost{} mitigation. Zeus~\cite{zeus} optimized energy for \emph{active} training. Patel et al.~\cite{patel-asplos24} characterized LLM inference power phases. Burtscher et al.~\cite{burtscher-power} showed \texttt{nvidia-smi} samples power for only $\sim$25\% of runtime---a caveat we address in \S\ref{sec:measurement}.

\paragraph{Concurrent work on execution-idle.} Lei et al.~\cite{lei-idle} characterize ``execution-idle''---intervals during job execution with near-zero GPU activity---via 31-day passive telemetry on a 756-GPU cluster spanning six GPU generations. They find 19.7\% of in-execution time (10.7\% of energy) is execution-idle, with serving workloads most affected (61\% of in-execution time). They propose two mitigations: automatic frequency downscaling (22--34\% power savings on L40S with latency tradeoffs) and load consolidation. They observe that GPU power and clock frequency remain elevated even after 2{,}048\,s of inactivity---corroborating our finding that the DVFS lock is indefinite once a CUDA context exists. Their work quantifies \emph{how often and at what cost} execution-idle occurs at cluster scale; ours isolates \emph{why} (CUDA context, not VRAM) via controlled per-GPU experiments across memory technologies. The parking tax we measure is the dominant sub-component of their execution-idle when GPU activity is strictly zero. Together, the two studies bracket the problem: theirs characterizes prevalence; ours identifies the mechanism and shows model size is irrelevant to idle power.

\paragraph{GPU sharing and multi-tenancy.} NVIDIA provides several mechanisms for sharing GPUs across workloads. Multi-Instance GPU (MIG)~\cite{nvidia-mig} partitions a GPU into isolated instances, each with its own CUDA context and memory. Multi-Process Service (MPS)~\cite{nvidia-mps} allows multiple CUDA contexts to share SM resources with reduced context-switching overhead. Time-slicing alternates CUDA contexts on shared hardware. All three approaches maintain at least one active CUDA context, meaning the DVFS overhead we measure persists. MIG partitions may amortize the overhead across tenants, but the aggregate GPU power is expected to remain similar, though per-tenant allocation of the overhead may differ. We do not evaluate multi-tenant configurations; whether MIG/MPS alter the per-context DVFS behavior is an open question.

\paragraph{Power capping.} NVIDIA's \texttt{nvidia-smi -pl} allows operators to set power limits below TDP, triggering clock throttling when the cap is approached. Whether power capping changes the idle DVFS step behavior---particularly whether a sufficiently low cap forces idle-with-context clocks below maximum boost---has not been studied. Our experiments use default power limits.

No prior work has measured the marginal idle power cost per GB of VRAM with formal equivalence bounds, decomposed idle power into DVFS and memory components across memory technologies, or evaluated vendor-provided mitigations for the boost-on-context behavior.

\section{Methodology}

\subsection{Phase~1: Production Telemetry}

\paragraph{Infrastructure.} We instrumented 14 NVIDIA H100 80GB SXM5 GPUs across 2~nodes running Kubernetes. NVIDIA Data Center GPU Manager (DCGM) Exporter exposes metrics to Prometheus at 30-second intervals.

\paragraph{Dataset.} We collected \num{336226} observations over 18~days. Average GPU utilization was 0.11\%; filtering to 0\% utilization retained \num{335267} samples (99.7\%). VRAM allocations ranged from 3\,MB to 77\,GB across five workload categories.

\paragraph{Effective sample size.} With 30\,s sampling and thermal correlation of 3--5~min ($\tau \approx 6$--$10$ samples):
\begin{equation}
    N_{\text{eff}} \approx \frac{N_{\text{raw}}}{2\tau + 1} \approx \num{16000}\text{--}\num{26000}
    \label{eq:neff}
\end{equation}

\subsection{Phase~2: Controlled Experiments}

We conducted within-subject dose-response experiments on four GPUs (Table~\ref{tab:gpus}). Each experiment followed the same protocol: record bare-idle baseline (no CUDA context), create a persistent CUDA context, then sequentially allocate increasing VRAM using PyTorch tensors (\texttt{torch.empty}). For each level: allocate, stabilize 60\,s, record \texttt{nvidia-smi} every 30\,s for 20~min, release, cool down 30\,s. The three primary GPUs (H100, A100, L40S) used this full 20-minute protocol; the B200 used a shorter confirmation sweep (3-minute phases, $n=18$). All GPUs were otherwise idle.

\begin{table}[tb]
\caption{Phase~2 GPU configurations. The first three GPUs used 30\,s sampling and 20-minute recording phases ($n=40$); the B200 used a shorter confirmation sweep (3-minute phases, $n=18$).}
\label{tab:gpus}
\small
\begin{tabular}{llrrr}
\toprule
\textbf{GPU} & \textbf{Memory} & \textbf{TDP} & \textbf{VRAM Range} & $n$/\textbf{phase} \\
\midrule
H100 80GB SXM & HBM3 & 700\,W & 0--64\,GB & 40 \\
A100 80GB PCIe & HBM2e & 300\,W & 0--72\,GB & 40 \\
L40S 48GB & GDDR6 & 350\,W & 0--40\,GB & 40 \\
B200 180GB SXM & HBM3e & 1000\,W & 0--32\,GB & 18 \\
\bottomrule
\end{tabular}
\end{table}

\subsection{Measurement Model}
\label{sec:measurement}

Within each treatment phase, we observe:
\begin{equation}
    \hat{P} = \frac{1}{n}\sum_{i=1}^{n} P(t_i), \quad \sigma_{\text{within}} < 1.5\,\text{W}
    \label{eq:sampling}
\end{equation}
The noisiest GPU (L40S, GDDR6) has $\sigma < 1.5$\,W; the cleanest (A100, HBM2e) has $\sigma < 0.08$\,W; the H100 (HBM3) has $\sigma < 0.17$\,W. With $n = 40$ samples per phase on the three primary architectures (H100, A100, L40S; the B200 used a shorter sweep with $n=18$), this yields SE $< 0.25$\,W on the noisiest device, ensuring sub-watt resolution for detecting systematic VRAM effects. On all four devices, the total power range across CUDA-active phases is $<$1\,W:
\begin{equation}
    \Delta P_{\text{VRAM}} < 1\,\text{W across full VRAM range on all GPUs}
    \label{eq:bound}
\end{equation}
This converts our measurement limitation into a \emph{result constraint}. Regarding Burtscher et al.'s~\cite{burtscher-power} finding that \texttt{nvidia-smi} undersamples: our within-phase standard deviations confirm idle power is stable enough that subsampling does not bias means. Complementary per-instruction energy models such as Wattchmen~\cite{wattchmen} achieve finer-grained decomposition; extending our idle-state analysis with such methods is a promising direction. For the shorter Blackwell and idle-recovery runs, and in the dose-response figures, we report the per-phase \emph{median} rather than the mean of Equation~\ref{eq:sampling}: the median is insensitive to the occasional transient DVFS boost a single sample can capture (one such sample appears in the Blackwell flag run), and on the 20-minute primary phases ($n=40$) the mean and median agree to within 1\,W.

\subsection{Phase~3: Vendor Mitigation Setup}
\label{sec:perf-boost-setup}

NVIDIA added the \perfboost{} environment variable in the 580 driver series (CUDA Toolkit 13.2, late 2025; publicly noted at driver 580.105.08), intended to prevent the idle-context clock boosting behavior documented by Patel et al.~\cite{patel-cal23} and confirmed in our Phase~1 and Phase~2 results. We evaluate its effectiveness on H100 80GB SXM (Hopper, driver 580.126.09) and A100 80GB SXM4 (Ampere, driver 580.126.16) under three conditions:

\begin{enumerate}
    \item \textbf{Baseline}: CUDA context + \texttt{torch.empty} 16\,GB, default driver behavior.
    \item \textbf{Flag on}: Same as (1) with \perfboost\texttt{=1} in the process environment.
    \item \textbf{Flag + vLLM}: vLLM serving Qwen2.5-7B~\cite{qwen2} in fp16, measuring idle power between requests with the flag on vs.\ off.
\end{enumerate}

Each condition follows the Phase~2 protocol: stabilize 60\,s, record \texttt{nvidia-smi} every 30\,s for the recording phase. For the vLLM conditions, we additionally measure request latency (p50, p95, p99) to detect performance regressions introduced by the flag.

\section{Results}

\subsection{Phase~1: Two Discrete Power States}

The production telemetry reveals a bimodal distribution:
\begin{itemize}
    \item \textbf{Bare idle} (345\,MHz): 7~GPUs, mean 74.7\,W $\pm$ 7.9\,W.
    \item \textbf{CUDA active} (1980\,MHz): 7~GPUs, mean 145.5\,W $\pm$ 11.2\,W.
\end{itemize}
The CUDA context effect is $+70.9$\,W (Cohen's $d = 7.3$, $p < 10^{-300}$). Within CUDA-active GPUs, regression across 3.2--77.1\,GB yields slope $= 0.013$\,W/GB ($R^2 = 0.001$, $p = 0.95$)---no detectable VRAM effect. The slope falls well within the $\pm 0.1$\,W/GB equivalence bound applied in Phase~2, consistent with VRAM independence at production scale. Node-level variation ($\sim$23\,W) exceeds any plausible VRAM effect, motivating Phase~2.

\subsection{Phase~2: Cross-Architecture Dose-Response}

\begin{figure*}[t]
    \centering
    \includegraphics[width=0.86\textwidth]{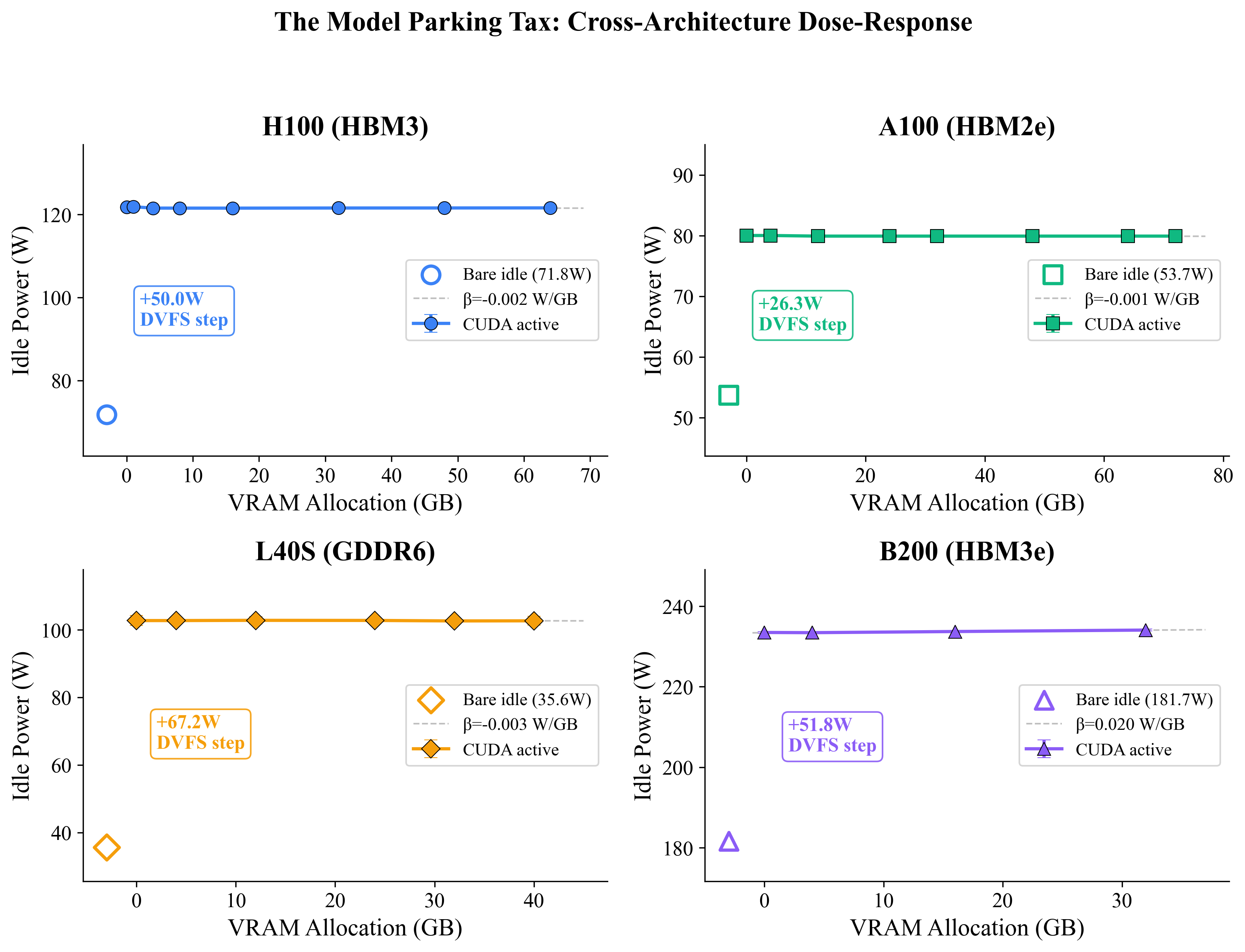}
    \caption{The model parking tax across four GPU architectures. Large markers (``bare''): bare-idle power with no CUDA context. Connected lines: CUDA-active phases with increasing VRAM. The discrete DVFS step dominates on every device, and all dose-response curves are flat across HBM3, HBM2e, GDDR6, and HBM3e (Blackwell). More model does not raise idle power.}
    \label{fig:hero}
\end{figure*}

DRAM physics predicts $\beta \approx 0$ (Equation~\ref{eq:mem-prediction}). This section tests that prediction and bounds its magnitude, so that the CUDA-context term is isolated as the sole determinant of idle power. Table~\ref{tab:cross} and Figure~\ref{fig:hero} present the controlled experiments across all four GPUs.

\begin{table}[tb]
\caption{Cross-architecture parking tax. Power values are per-phase means (medians for B200; \S\ref{sec:measurement}). $\beta$ is the linear regression slope across CUDA-active phases. All GPUs at 0\% utilization.}
\label{tab:cross}
\scriptsize
\setlength{\tabcolsep}{3.5pt}
\begin{tabular}{lrrrr}
\toprule
 & \textbf{H100} & \textbf{A100} & \textbf{L40S} & \textbf{B200} \\
 & \textbf{(HBM3)} & \textbf{(HBM2e)} & \textbf{(GDDR6)} & \textbf{(HBM3e)} \\
\midrule
Bare idle (W) & 71.8 & 53.7 & 35.6 & 181.7 \\
SM clk, bare (MHz) & 345 & 210 & 210 & 120 \\
CUDA ctx power (W) & 121.7 & 80.0 & 102.7 & 233.6 \\
SM clk, ctx (MHz) & 1980 & 1410 & 2520 & 1965 \\
\midrule
\textbf{Context overhead (W)} & \textbf{+49.9} & \textbf{+26.3} & \textbf{+67.2} & \textbf{+51.9} \\
Context (\% of TDP) & 7.1\% & 8.8\% & 19.2\% & 5.2\% \\
\midrule
Max VRAM tested (GB) & 64 & 72 & 40 & 32 \\
Power range, VRAM (W) & +0.34 & $-$0.09 & +0.08 & +0.60 \\
$\beta$ (W/GB) & $-$0.002 & $-$0.001 & $-$0.002 & $+$0.019 \\
$p$-value ($\beta = 0$) & 0.404 & 0.002$^\dagger$ & 0.794 & $<$0.001$^\dagger$ \\
$p_{\text{TOST}}$ ($|\beta| < 0.1$) & $10^{-9}$ & $<0.001$ & $<0.05$ & $<0.01$ \\
\midrule
\textbf{Context \% of tax} & \textbf{$>$99\%} & \textbf{$>$99\%} & \textbf{$>$99\%} & \textbf{$>$99\%} \\
\bottomrule
\end{tabular}
\par\vspace{2pt}
\parbox{\columnwidth}{\footnotesize $^\dagger$Slope significant but a temporal/thermal artifact of the sequential sweep (A100 negative, B200 positive), $<$1\,W total and far below the 0.1\,W/GB bound; see \S\ref{sec:a100-note}. B200 values are per-phase medians (transient-robust); others are means, agreeing within 1\,W.}
\end{table}

\paragraph{CUDA context overhead.} On all four GPUs, creating a CUDA context triggers a discrete DVFS transition: SM clocks jump from 120--345\,MHz to 1410--2520\,MHz (architecture-dependent). The associated power step ranges from $+26.3$\,W (A100) to $+67.2$\,W (L40S). This is the dominant component of the parking tax on every architecture.

\begin{figure*}[t]
    \centering
    \includegraphics[width=0.86\textwidth]{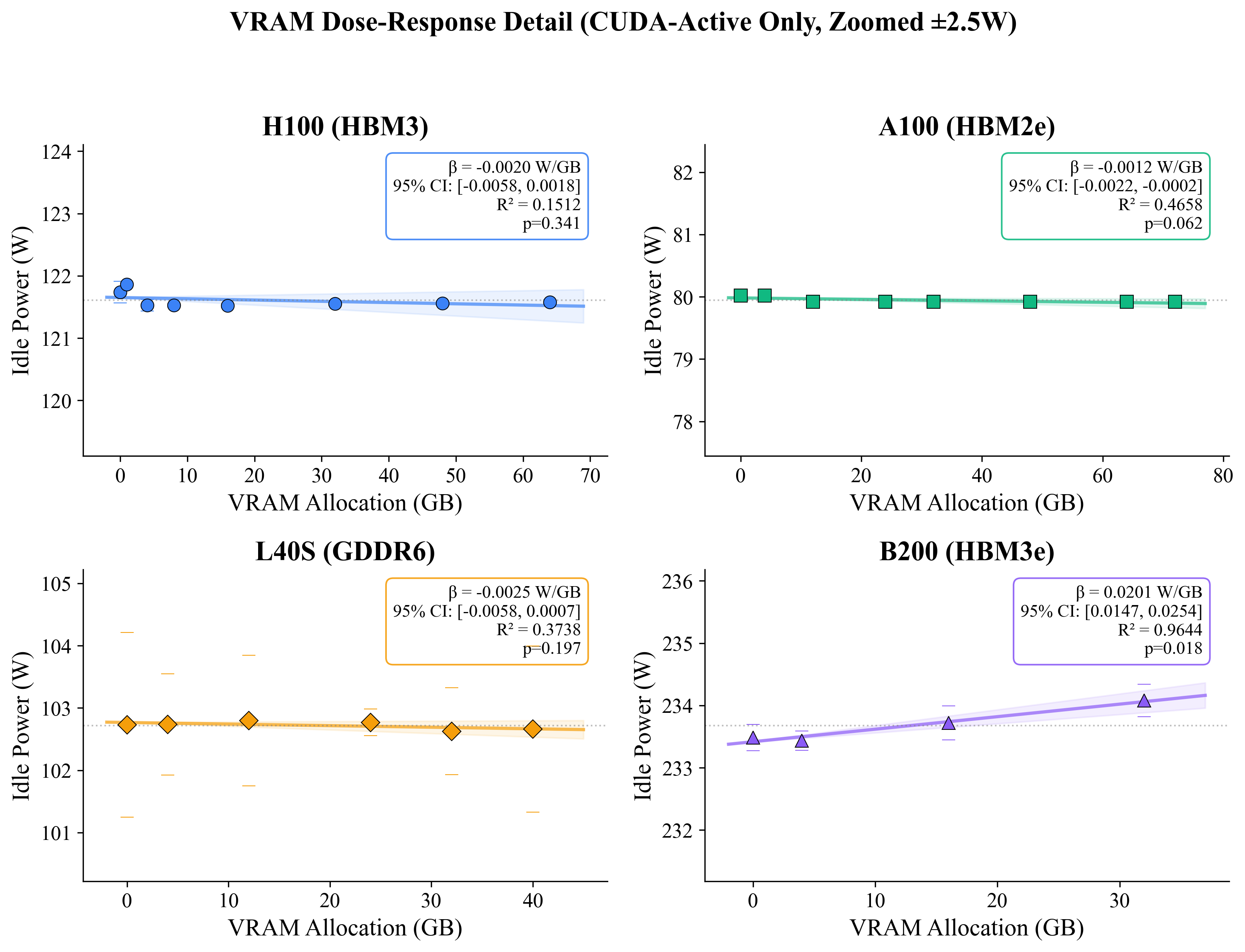}
    \caption{VRAM dose-response detail (CUDA-active phases only, zoomed to $\pm$2.5\,W). Each panel shows one GPU with regression line. H100 and L40S slopes are not significant; the A100 (negative) and B200 (positive) slopes are significant but are tiny temporal/thermal artifacts of the sequential sweep, not memory effects. All four are flat within $\pm$1\,W.}
    \label{fig:regression}
\end{figure*}

\paragraph{VRAM allocation: bounded below relevance.} Once a CUDA context is active, varying VRAM from 0 to 40--72\,GB changes power by less than 1\,W on \emph{every GPU tested}. The power range reflects measurement noise; the regression slope $\beta$ captures any systematic effect. Per-GPU 95\% confidence intervals for $\beta$:
\begin{itemize}
    \item H100: $\beta = -0.002$\,W/GB, 95\% CI $[-0.005, +0.002]$, $p = 0.40$
    \item A100: $\beta = -0.001$\,W/GB, 95\% CI $[-0.002, -0.0005]$, $p = 0.002^\dagger$
    \item L40S: $\beta = -0.002$\,W/GB, 95\% CI $[-0.019, +0.015]$, $p = 0.79$
\end{itemize}
We fail to reject $\beta = 0$ on the H100 and L40S. On the A100, $\beta$ is statistically significant but negative (see below). To move beyond the absence of evidence, we apply the Two One-Sided Tests (TOST) equivalence procedure~\cite{schuirmann-tost} with an equivalence bound of $\Delta = \pm 0.1$\,W/GB---chosen so that even a 64\,GB model would contribute $<$6.4\,W---several-fold smaller than the DVFS overhead (26--67\,W) on every architecture tested. This ensures the equivalence bound is conservative: any effect below $\Delta$ is operationally irrelevant for keep-warm decisions. On all four GPUs, the TOST procedure rejects the hypothesis that $|\beta| \geq 0.1$\,W/GB ($p_{\text{TOST}} < 0.05$): under the tested protocol we do not merely fail to find a memory effect, but bound it several-fold below the DVFS step. Two caveats attend this bound. The sweep is sequential on one device per architecture, so the within-phase samples are correlated and the effective observations are the per-phase means rather than the 40 raw samples. And the sequential design cannot fully separate a small VRAM term from slow time or thermal drift---the A100 and B200 slopes are two such confounds. Figure~\ref{fig:regression} shows the zoomed regression detail for all four GPUs.

\paragraph{The A100 negative slope.}
\label{sec:a100-note}
On the A100, $\beta = -0.001$\,W/GB with $p = 0.002$---statistically significant but \emph{negative}. The total effect is $-0.09$\,W across 72\,GB, coinciding with a 0.7\textdegree C HBM temperature drift (50.2\textdegree C $\to$ 49.5\textdegree C) over the sequential experiment. This is a temporal confound, not a memory physics effect: more VRAM cannot \emph{reduce} power. It demonstrates that even when $p < 0.05$, the detected signal is 300$\times$ smaller than the DVFS component and directionally implausible.

\begin{figure*}[t]
    \centering
    \includegraphics[width=0.86\textwidth]{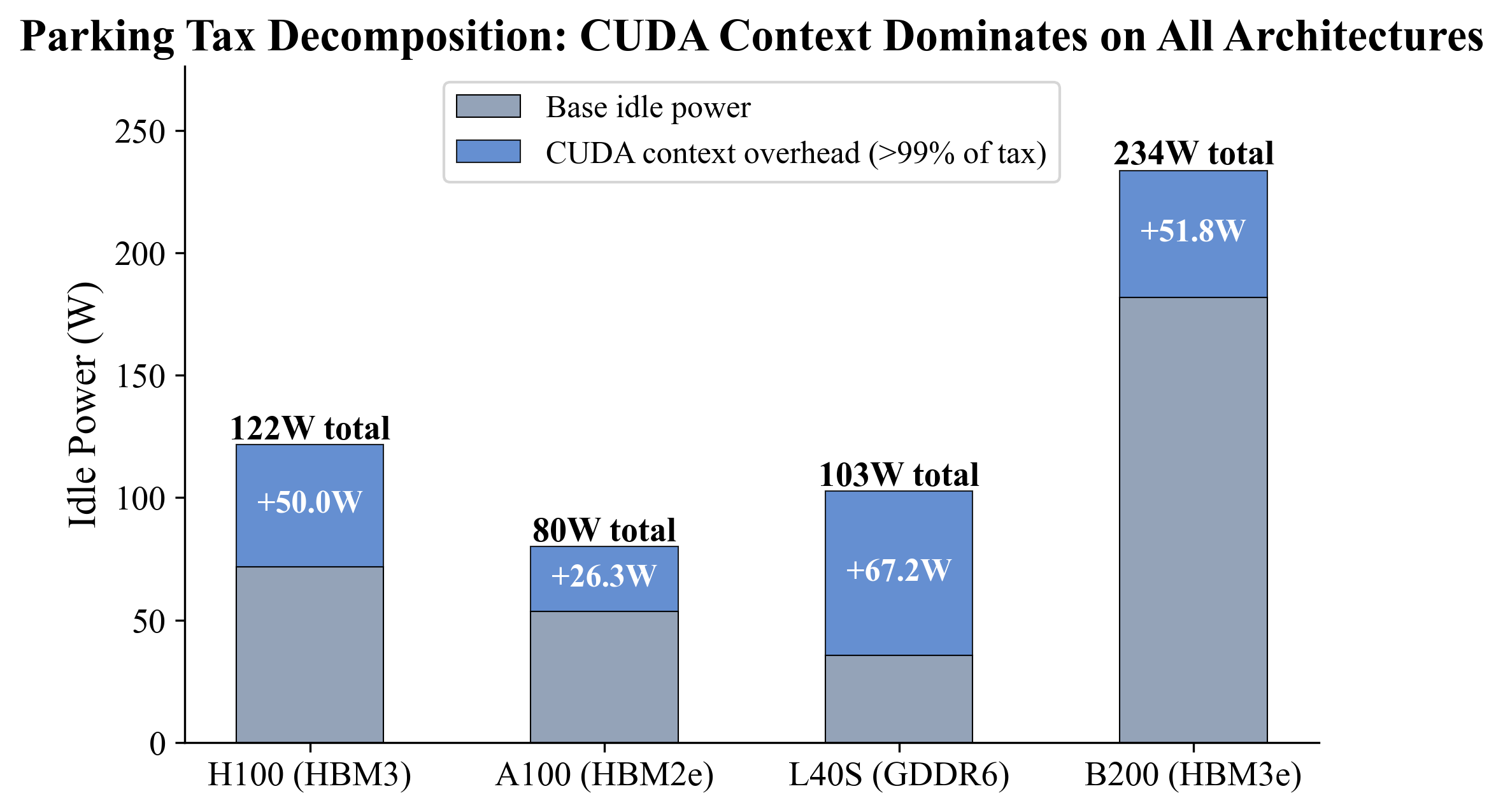}
    \caption{Parking tax decomposition. Gray: base idle power. Color: CUDA context overhead ($>$99\% of the tax on every architecture). The VRAM term is too small to render.}
    \label{fig:decomp}
\end{figure*}

\paragraph{System characterization.} The idle power model (Equation~\ref{eq:model}) is empirically, per architecture:
\begin{align}
    P_{\text{H100}} &\approx 71.8 + 49.9 \cdot \mathbf{1}_{[C=1]} \;\text{W} \label{eq:h100}\\
    P_{\text{A100}} &\approx 53.7 + 26.3 \cdot \mathbf{1}_{[C=1]} \;\text{W} \label{eq:a100}\\
    P_{\text{L40S}} &\approx 35.6 + 67.2 \cdot \mathbf{1}_{[C=1]} \;\text{W} \label{eq:l40s}
\end{align}
with $\beta \approx 0$ on all four. The parking tax is a step function determined by CUDA context presence, invariant to memory occupancy and memory technology. Figure~\ref{fig:decomp} visualizes this decomposition: the VRAM component is invisible at scale on all four architectures.

\paragraph{Cross-technology consistency.} The flat dose-response holds across HBM3, HBM2e, and GDDR6---three fundamentally different memory subsystems with different bandwidths (3.35\,TB/s, 2.0\,TB/s, 864\,GB/s), clock rates (2619, 1512, 9001\,MHz), and physical packaging. This is consistent with the memory physics prediction of Equation~\ref{eq:mem-prediction}: at idle, $\partial P_{\text{mem}} / \partial \text{occupancy} \approx 0$ regardless of DRAM technology.

\begin{figure}[t]
    \centering
    \includegraphics[width=\columnwidth]{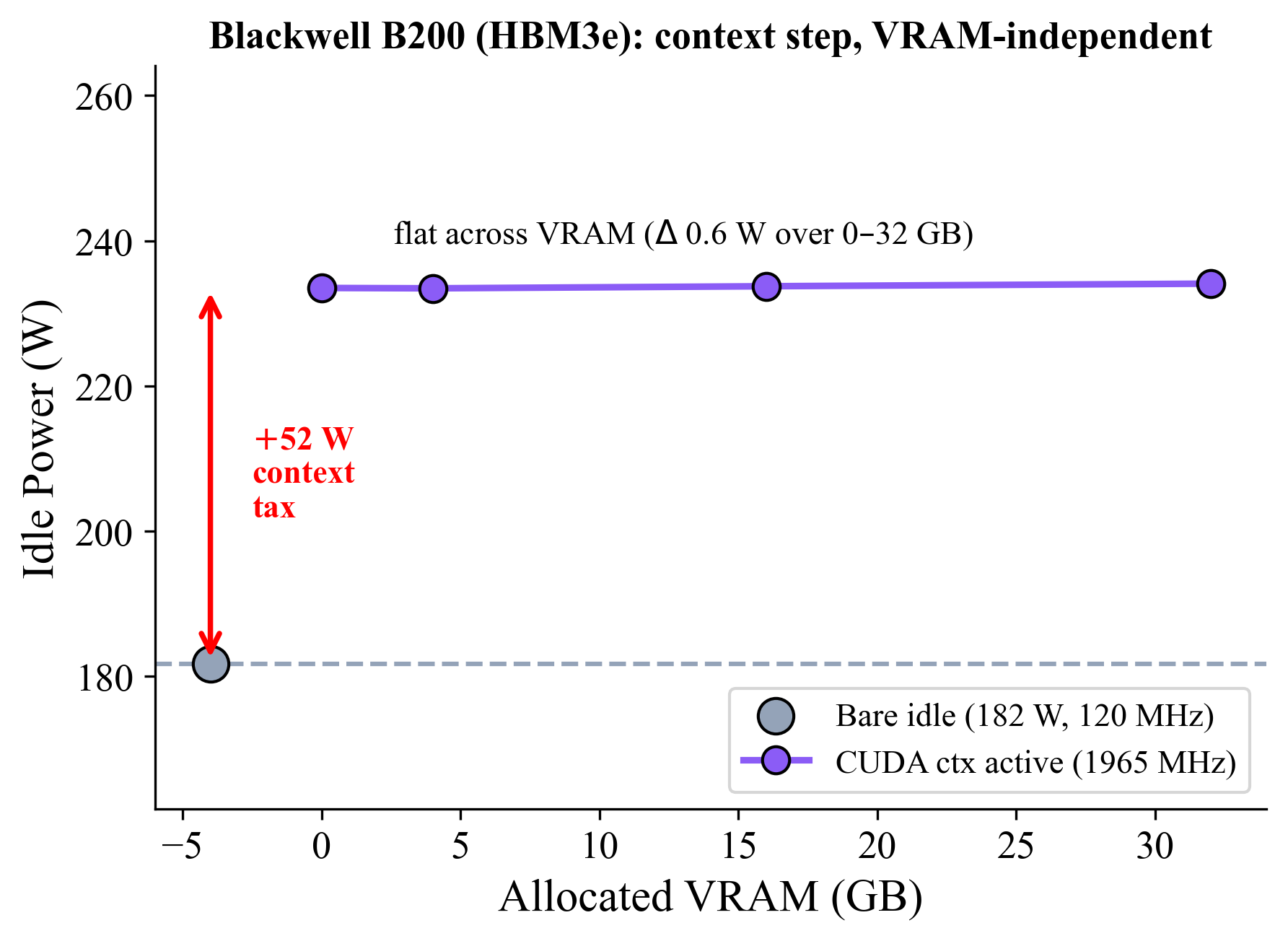}
    \caption{Dose-response on an NVIDIA B200 (Blackwell, HBM3e); four VRAM levels, $n=18$ per phase. Creating a CUDA context adds a 52\,W tax (182$\to$234\,W) at 0\% utilization; power then stays flat within 1\,W across 0--32\,GB, matching the other three architectures.}
    \label{fig:b200}
\end{figure}

\paragraph{Blackwell (HBM3e).} On the B200 (Blackwell, HBM3e) we ran the same dose-response over four VRAM levels (3-minute phases, $n=18$; Figure~\ref{fig:b200}). Creating a context raises idle power from 182\,W (120\,MHz) to 234\,W (1965\,MHz)---a 52\,W tax at 0\% utilization---and power stays flat within 1\,W across 0--32\,GB, matching the other three architectures. The boosting default is present on this current-generation part.

\subsection{Real Model Validation}

The Phase~2 experiments use \texttt{torch.empty} tensors to isolate the VRAM effect from model-specific CUDA state. To confirm that real model weights---with their CUDA kernels, cuDNN handles, and weight tensors---do not change the result, we loaded a production Qwen2.5-7B model (fp16~\cite{qwen2}) and measured idle power (Table~\ref{tab:validation}).

\begin{table}[tb]
\caption{Real-model validation on the H100. A real Qwen2.5-7B (fp16) idles within 0.03\,W of a 16\,GB \texttt{torch.empty} allocation holding the same CUDA context; the delta is within measurement noise.}
\label{tab:validation}
\small
\begin{tabular}{@{}llrrr@{}}
\toprule
\textbf{GPU} & \textbf{Allocation} & \textbf{VRAM (GB)} & \textbf{Idle (W)} & $\Delta$ \\
\midrule
\multirow{2}{*}{H100} & \texttt{torch.empty} 16\,GB & 16.0 & 121.51 & --- \\
 & Qwen2.5-7B (fp16) & 14.9 & 121.48 & $-$0.03\,W \\
\bottomrule
\end{tabular}
\end{table}

On the H100, Qwen2.5-7B idles at 121.48\,W---within 0.03\,W of the \texttt{torch.empty} baseline ($n=30$, $\sigma = 0.11$\,W), and far below the 26--67\,W CUDA context overhead. A real 7B model loaded on the L40S idles at 105.85\,W, in line with the context-active idle level measured for that architecture in Phase~2; absolute idle power varies across GPU instances, tracking the $\sim$23\,W inter-node spread observed in Phase~1, while the within-instance VRAM invariance holds regardless. Whether a GPU holds a real 7B model or an equivalent block of empty memory, the idle cost is set by whether a context is resident, not by the model.

\paragraph{Cold-start power profile.} On the H100, we captured 1\,Hz power traces during Qwen2.5-7B loading (29.7\,s). The profile is bursty: 22\,s at bare idle ($\sim$70.8\,W, CPU-side deserialization), a 3\,s GPU burst peaking at 124.1\,W during weight transfer to VRAM, then settling to the CUDA-active idle ($\sim$121\,W). The peak loading power (124\,W) is far below TDP (700\,W), consistent with the I/O-bound nature of model loading. On the A100 and L40S, Qwen2.5-7B loaded in 11.5\,s and 10.8\,s respectively (both fp16, $\sim$14.8\,GB VRAM), corroborating that cold starts for small models complete in seconds across architectures. These measured profiles validate using $P_{\text{load}} \approx 100$--$300$\,W in our breakeven model (Equation~\ref{eq:breakeven}).

\subsection{Cold-Start Breakeven Model}

Given a parking tax of $P_{\text{park}}$ watts, an operator may instead evict the model and cold-start it on the next request. The energy breakeven between keeping warm and cold-starting is:
\begin{equation}
    T^* = \frac{P_{\text{load}} \cdot t_{\text{load}}}{P_{\text{park}}}
    \label{eq:breakeven}
\end{equation}
where $P_{\text{load}}$ is the mean power during model loading and $t_{\text{load}}$ is the loading duration.

Table~\ref{tab:breakeven} shows breakeven intervals for the H100 ($P_{\text{park}} = 49.9$\,W). For standard PyTorch loading of a 70B model (45\,s, $\sim$300\,W), $T^* \approx$ \textbf{4.5~min}. On the A100 ($P_{\text{park}} = 26.3$\,W), $T^*$ is longer ($\sim$8.5~min) because the lower overhead takes longer to exceed the cold-start cost. On the L40S ($P_{\text{park}} = 67.2$\,W), $T^*$ is shorter ($\sim$3.4~min).

\begin{table}[tb]
\caption{Cold-start energy breakeven on H100 ($P_{\text{park}} = 49.9$\,W). $^\dagger$Measured in this work; others estimated from published benchmarks.}
\label{tab:breakeven}
\small
\begin{tabular}{lrrr}
\toprule
\textbf{Loading Method} & $P_{\text{load}}$ \textbf{(W)} & $t_{\text{load}}$ \textbf{(s)} & $T^*$ \\
\midrule
Qwen2.5-7B (measured)$^\dagger$ & 124 & 30 & 1.2\,min \\
Standard PyTorch (70B) & 300 & 45 & 4.5\,min \\
ServerlessLLM (70B) & 300 & 8 & 48\,s \\
Run:ai Streamer (8B) & 200 & 5 & 20\,s \\
\bottomrule
\end{tabular}
\end{table}

\paragraph{Queueing extension.} If requests arrive as a Poisson process with rate $\lambda$, the optimal eviction threshold is $\lambda^* = P_{\text{park}} / (P_{\text{load}} \cdot t_{\text{load}})$. For H100 with PyTorch loading: $\lambda^* \approx 13$\,req/hr; below this rate, cold-starting saves energy.

\paragraph{A third option.} The breakeven analysis assumes a binary keep-warm-or-evict decision. A third option keeps the model loaded while removing the parking tax at the driver level, paying only a transient ramp cost per idle-to-active transition instead of a full cold start. We evaluate it next.

\subsection{Phase~3: Vendor Mitigation}
\label{sec:perf-boost}

NVIDIA's \perfboost{} environment variable (added in the 580 driver series; tested at $\geq$580.105.08, November 2025) is the most direct vendor-provided mechanism targeting the boost-on-context behavior. Using the experimental protocol described in \S\ref{sec:perf-boost-setup}, we evaluate its effectiveness on H100 (Hopper) and A100 (Ampere).

\begin{table}[tb]
\caption{Effect of \perfboost{} on idle-with-context power. H100/A100: mean $\pm$ std ($n=40$, 20-min phases; drivers 580.126.09 and 580.126.16). L40S and B200: shorter idle-only protocol (per-phase medians).}
\label{tab:perf-boost}
\small
\setlength{\tabcolsep}{3.5pt}
\begin{tabular}{@{}llrrrr@{}}
\toprule
\textbf{GPU} & \textbf{Condition} & \textbf{Power (W)} & \textbf{SM (MHz)} & \textbf{PState} & $\Delta$\textbf{P} \\
\midrule
\multirow{3}{*}{\shortstack[l]{H100 SXM\\(Hopper)}} & Bare idle & 73.3 $\pm$ 0.1 & 345 & P0 & --- \\
 & Baseline & 128.1 $\pm$ 0.4 & 1980 & P0 & +54.8 \\
 & Flag on & 73.2 $\pm$ 0.1 & 345 & P0 & $-$54.9 \\
\midrule
\multirow{3}{*}{\shortstack[l]{A100 SXM4\\(Ampere)}} & Bare idle & 64.0 $\pm$ 0.3 & 210 & --- & --- \\
 & Baseline & 75.1 $\pm$ 0.5 & 1155 & --- & +11.1 \\
 & Flag on & 64.5 $\pm$ 0.4 & 210 & --- & $-$10.6 \\
\midrule
\multirow{3}{*}{\shortstack[l]{L40S\\(Ada Lovelace)}} & Bare idle & 35.6 $\pm$ 0.1 & 210 & --- & --- \\
 & Baseline & 102.7 $\pm$ 0.7 & 2520 & --- & +67.1 \\
 & Flag on & 36.7 $\pm$ 0.1 & 210 & --- & $-$66.0 \\
\midrule
\multirow{3}{*}{\shortstack[l]{B200\\(Blackwell)}} & Bare idle & 181.7 $\pm$ 0.2 & 120 & --- & --- \\
 & Baseline & 233.5 $\pm$ 0.2 & 1965 & --- & +51.8 \\
 & Flag on & 181.8 $\pm$ 0.3 & 120 & --- & $-$51.7 \\
\bottomrule
\end{tabular}
\end{table}

\begin{table}[tb]
\caption{vLLM idle power with \perfboost. Qwen2.5-7B (fp16) served via vLLM, idle between requests ($n=40$, 20-min phases).}
\label{tab:vllm-idle}
\small
\begin{tabular}{@{}llrrr@{}}
\toprule
\textbf{GPU} & \textbf{Condition} & \textbf{Power (W)} & \textbf{SM (MHz)} & \textbf{VRAM (GB)} \\
\midrule
\multirow{2}{*}{H100 SXM} & vLLM baseline & 128.6 $\pm$ 1.2 & 1980 & 72.5 \\
 & vLLM flag on & 73.1 $\pm$ 0.2 & 345 & 72.7 \\
\midrule
\multirow{2}{*}{A100 SXM4} & vLLM baseline & 73.9 $\pm$ 0.2 & 1155 & 72.3 \\
 & vLLM flag on & 64.3 $\pm$ 0.4 & 210 & 72.3 \\
\bottomrule
\end{tabular}
\end{table}

\begin{table}[tb]
\caption{Steady-state vLLM inference latency with and without \perfboost. Qwen2.5-7B (fp16), $n=50$ warm requests per condition (after 3 warmup requests and 60\,s idle soak), 128-token prompt / 128-token completion.}
\label{tab:latency}
\small
\setlength{\tabcolsep}{3.5pt}
\begin{tabular}{@{}llrrrr@{}}
\toprule
\textbf{GPU} & \textbf{Condition} & \textbf{Mean (ms)} & \textbf{p50 (ms)} & \textbf{p95 (ms)} & \textbf{p99 (ms)} \\
\midrule
\multirow{2}{*}{H100 SXM} & Baseline & 533.4 & 533.5 & 534.7 & 535.2 \\
 & Flag on & 533.3 & 533.2 & 534.9 & 536.4 \\
\midrule
\multirow{2}{*}{A100 SXM4} & Baseline & 910.0 & 910.0 & 910.9 & 911.5 \\
 & Flag on & 910.0 & 910.1 & 911.5 & 912.0 \\
\bottomrule
\end{tabular}
\end{table}

\begin{table}[tb]
\caption{First-request DVFS ramp after 60\,s idle soak. With the flag, SM clocks are at bare-idle frequency; the first inference request triggers a clock ramp. By request~2, latency returns to steady state.}
\label{tab:cold-ramp}
\small
\setlength{\tabcolsep}{3pt}
\begin{tabular}{@{}llrrr@{}}
\toprule
\textbf{GPU} & \textbf{Condition} & \textbf{Req~1 (ms)} & \textbf{Req~2--5 avg (ms)} & \textbf{Penalty} \\
\midrule
\multirow{2}{*}{H100 SXM} & Baseline & 546 & 536 & +10\,ms \\
 & Flag on & 683 & 536 & +147\,ms \\
\midrule
\multirow{2}{*}{A100 SXM4} & Baseline & 921 & 910 & +11\,ms \\
 & Flag on & 1{,}130 & 910 & +220\,ms \\
\bottomrule
\end{tabular}
\end{table}

\paragraph{H100 results.} Table~\ref{tab:perf-boost} summarizes the flag's effect on idle power. On the H100 (driver 580.126.09, $n=40$, 20-min phases), \perfboost\texttt{=1} eliminates the parking tax within measurement noise. With the flag set, the SM clock remains at the bare-idle frequency of 345\,MHz---identical to the no-CUDA-context state---even though a CUDA context is active and 16\,GB of VRAM is allocated. Idle power drops from 128.1\,W (baseline) to 73.2\,W (median; two transient DVFS samples raise the mean to 75.9\,W), a reduction of 54.9\,W that accounts for $>$100\% of the DVFS overhead measured in Phase~2 ($\Delta P_{\text{DVFS}} = 49.9$\,W in Table~\ref{tab:cross}; the overshoot is within inter-session variance). The flag-on power (73.2\,W median) is indistinguishable from bare idle (73.3 $\pm$ 0.1\,W), confirming that the entire parking tax is attributable to the DVFS boost and that the flag successfully suppresses it. The effect persists through the full 20-minute recording period with no sign of driver heuristics overriding the setting (38/40 samples clean; 2 exhibited transient spikes consistent with driver telemetry events).

\paragraph{A100 results.} On the A100 SXM4 (driver 580.126.16, $n=40$, 20-min phases), the flag produces the same structural result. Without the flag, a CUDA context boosts the SM clock from 210\,MHz to 1155\,MHz, adding 11.1\,W over bare idle. With the flag set, the SM clock remains at 210\,MHz and idle power drops from 75.1\,W (baseline) to 64.5\,W---within 0.5\,W of bare idle (64.0\,W). The 10.6\,W reduction recovers 96\% of the DVFS overhead (10.6 of 11.1\,W; the 0.5\,W residual is within measurement noise). This A100 SXM4 variant boosts to a lower idle clock (1155\,MHz) than the PCIe variant tested in Phase~2 (1410\,MHz, 26.3\,W overhead), but the flag eliminates the overhead on both form factors, confirming the mitigation is not variant-specific.

\begin{figure*}[t]
    \centering
    \includegraphics[width=0.86\textwidth]{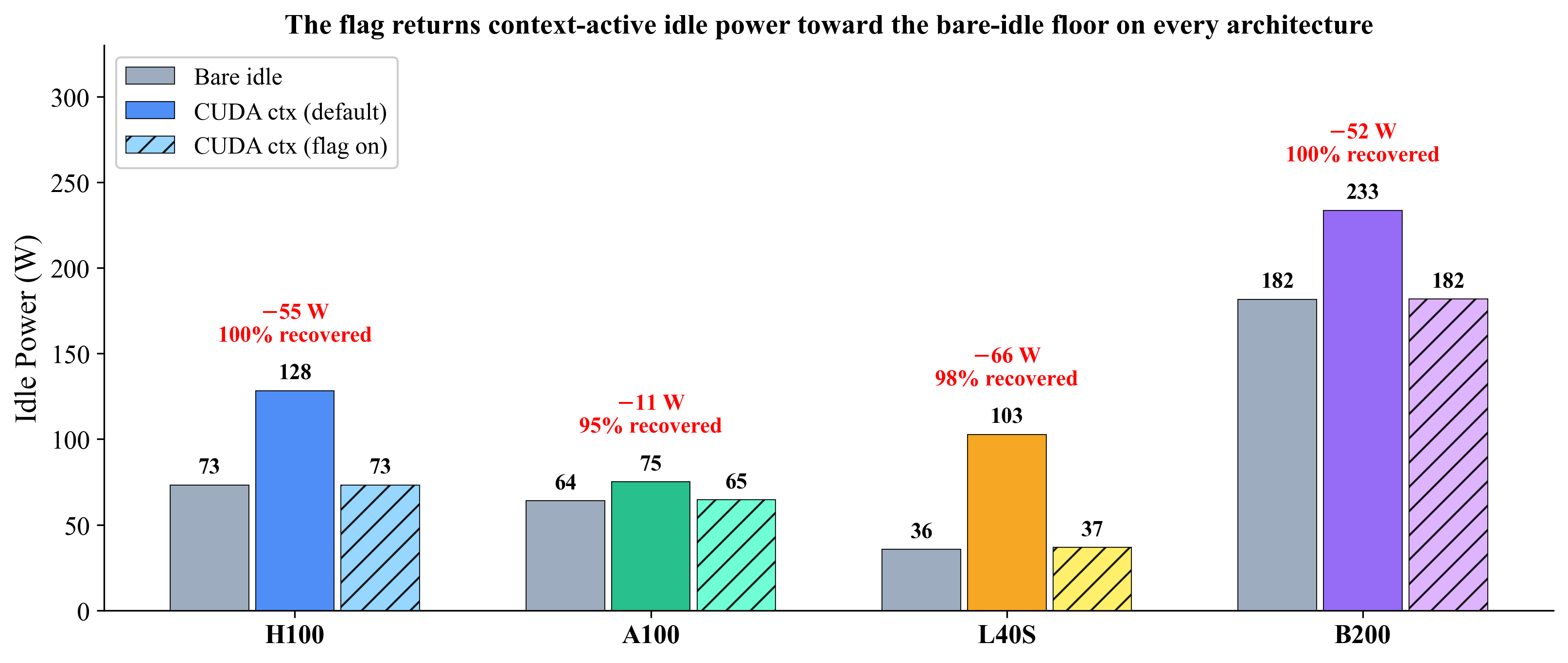}
    \caption{The \perfboost{} flag returns context-active idle power toward the bare-idle floor on all four architectures (H100, A100, L40S, B200). Recovery is roughly 95--100\% on all four; on the B200 the flag returns context-active idle to the 120\,MHz bare floor (234$\to$182\,W).}
    \label{fig:flag-recovery}
\end{figure*}

\paragraph{The mitigation holds across four architectures.} Beyond the H100 and A100 (Table~\ref{tab:perf-boost}), we measured the flag's idle-power effect on the L40S and B200 with an idle-only protocol (Figure~\ref{fig:flag-recovery}). On the L40S---the highest-tax device---context-active idle falls from 102\,W (2520\,MHz) to 36.8\,W (210\,MHz), the bare-idle floor: essentially the full tax recovered. On the Blackwell B200 the flag returns idle power to the same floor: with the flag set, median context-active idle falls from 234\,W to 182\,W (the bare-idle floor), and the SM clock dropped to 120\,MHz in 71 of 72 context-active samples, with one transient boost during the 4\,GB phase. Across all four architectures the flag recovers roughly 95--100\% of the context overhead.

\begin{figure*}[t]
    \centering
    \includegraphics[width=0.86\textwidth]{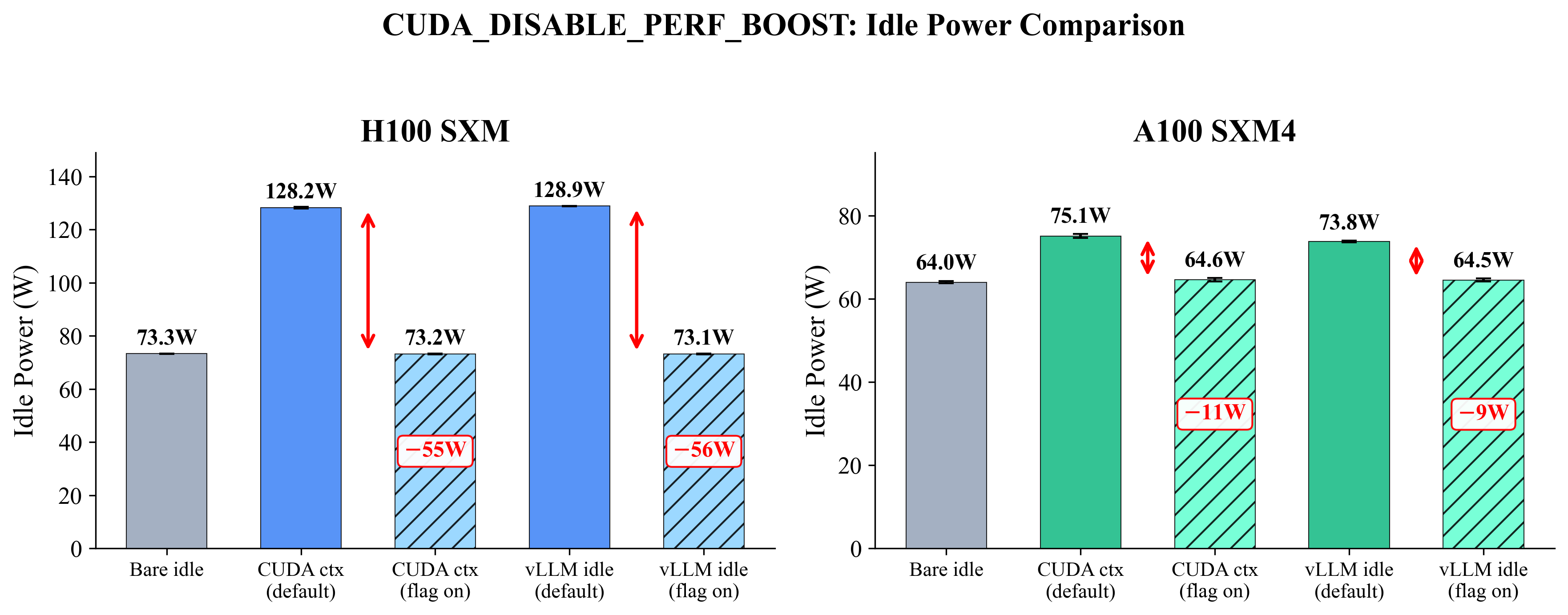}
    \caption{Effect of \perfboost{} on idle power. Hatched bars indicate the flag is enabled. On both H100 and A100, the flag reduces idle-with-context power to bare-idle levels---whether the CUDA context comes from \texttt{torch.empty} or a full vLLM serving stack with $\sim$72\,GB of model state.}
    \label{fig:perf-boost-power}
\end{figure*}

\paragraph{vLLM idle power.} To validate through a real serving stack, we measured vLLM~\cite{vllm} idle power serving Qwen2.5-7B (fp16) on both architectures (Table~\ref{tab:vllm-idle}, Figure~\ref{fig:perf-boost-power}). On H100, vLLM baseline idles at 128.6\,W; with the flag, it drops to 73.1\,W ($-$55.5\,W). On A100, vLLM baseline idles at 73.9\,W; with the flag, 64.3\,W ($-$9.6\,W). VRAM usage is unchanged ($\sim$72\,GB on both). The flag's effect is identical whether the CUDA context comes from \texttt{torch.empty} or a full serving framework.

\paragraph{Inference latency.} We test whether the flag adds latency when requests arrive and the GPU must ramp clocks from idle, using a focused retest protocol: start vLLM, send 3 warmup requests, idle for 60\,s (allowing clocks to drop under the flag), then send 5 ``cold'' requests followed by 50 ``warm'' requests back-to-back.

\begin{figure*}[t]
    \centering
    \includegraphics[width=0.86\textwidth]{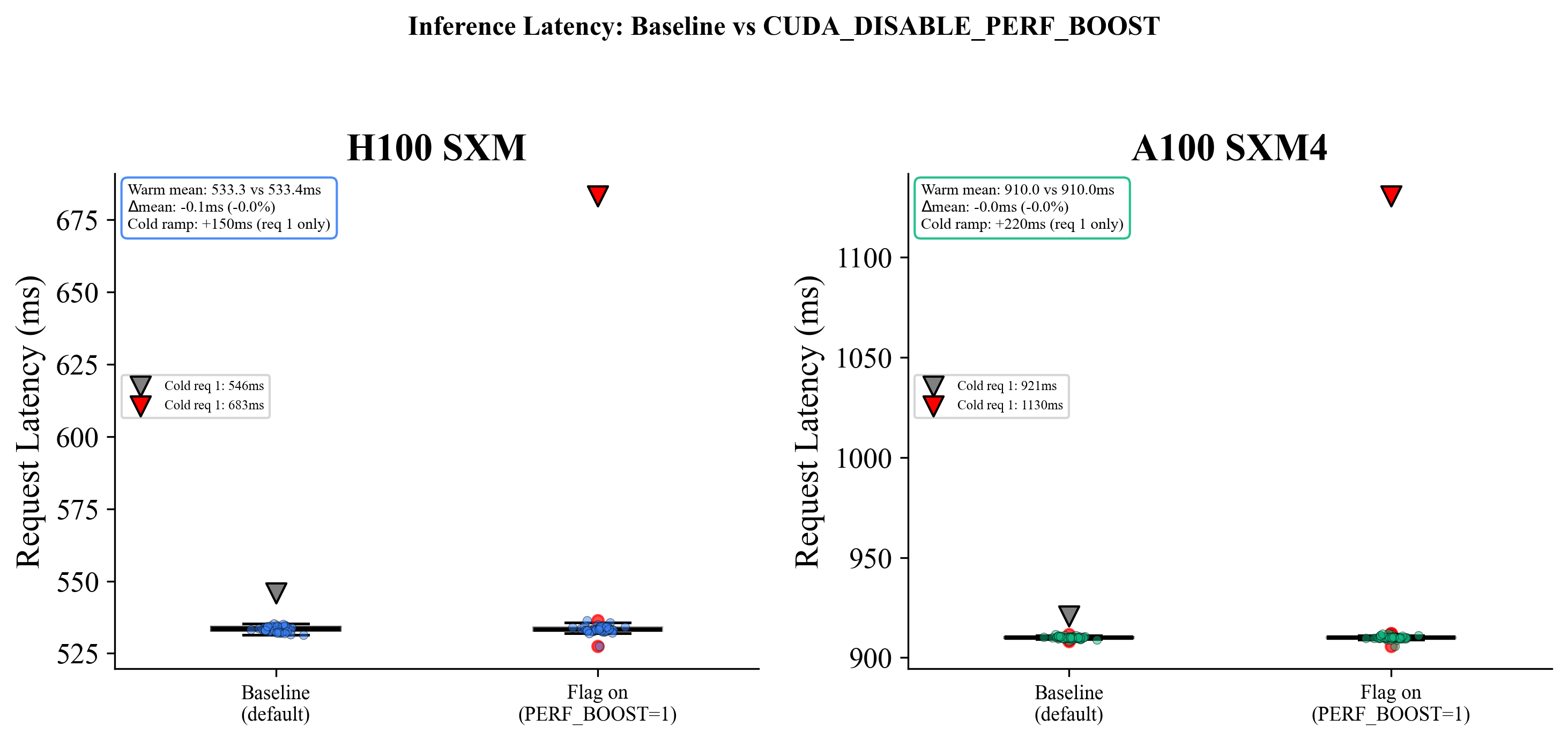}
    \caption{Inference latency with and without \perfboost. Box plots show $n=50$ warm (steady-state) requests; triangles mark the first request after 60\,s idle. Warm distributions are identical on both GPUs. The cold first request (red triangle) shows the DVFS ramp penalty: $+$147\,ms on H100, $+$220\,ms on A100---gone by request~2.}
    \label{fig:perf-boost-latency}
\end{figure*}

\emph{Steady state} (Table~\ref{tab:latency}, Figure~\ref{fig:perf-boost-latency}): Across 50 warm requests on both GPUs, latency is identical with and without the flag. On the H100, mean latency is 533.3\,ms (flag on) vs.\ 533.4\,ms (baseline); on the A100, 910.0\,ms vs.\ 910.0\,ms. All percentiles through p99 are within 1.2\,ms on both architectures. vLLM uses CUDA graphs for decode-phase execution; the identical latency distributions confirm that CUDA graphs function correctly under the flag.

\emph{First-request ramp} (Table~\ref{tab:cold-ramp}): The flag does introduce a one-time DVFS ramp penalty on the first request after an idle period. On the H100, the first request after 60\,s idle takes 683\,ms with the flag ($+$147\,ms over steady state) vs.\ 546\,ms without ($+$10\,ms). On the A100, the penalty is 1{,}130\,ms ($+$220\,ms) vs.\ 921\,ms ($+$11\,ms). By request~2, latency returns to steady state on both architectures. The ramp is a one-time cost per idle-to-active transition, mitigable with a single warmup request or keep-alive probe. This contrasts with Lei et al.'s~\cite{lei-idle} frequency downscaling, which achieves 22--34\% idle power savings on L40S but with \emph{persistent} latency tradeoffs throughout active inference. The flag eliminates the parking tax during idle while preserving full clock speeds during sustained inference---comparable savings with only a transient cost at the idle-to-active boundary.

This reframes the breakeven model (\S4.4): instead of choosing between keep-warm (pay the tax) and evict (pay the cold start), the flag offers a third option---keep-warm with no tax and only a ${\sim}150$\,ms ramp. For workloads that must remain responsive, this appears to dominate both alternatives.

\begin{figure*}[t]
    \centering
    \includegraphics[width=0.86\textwidth]{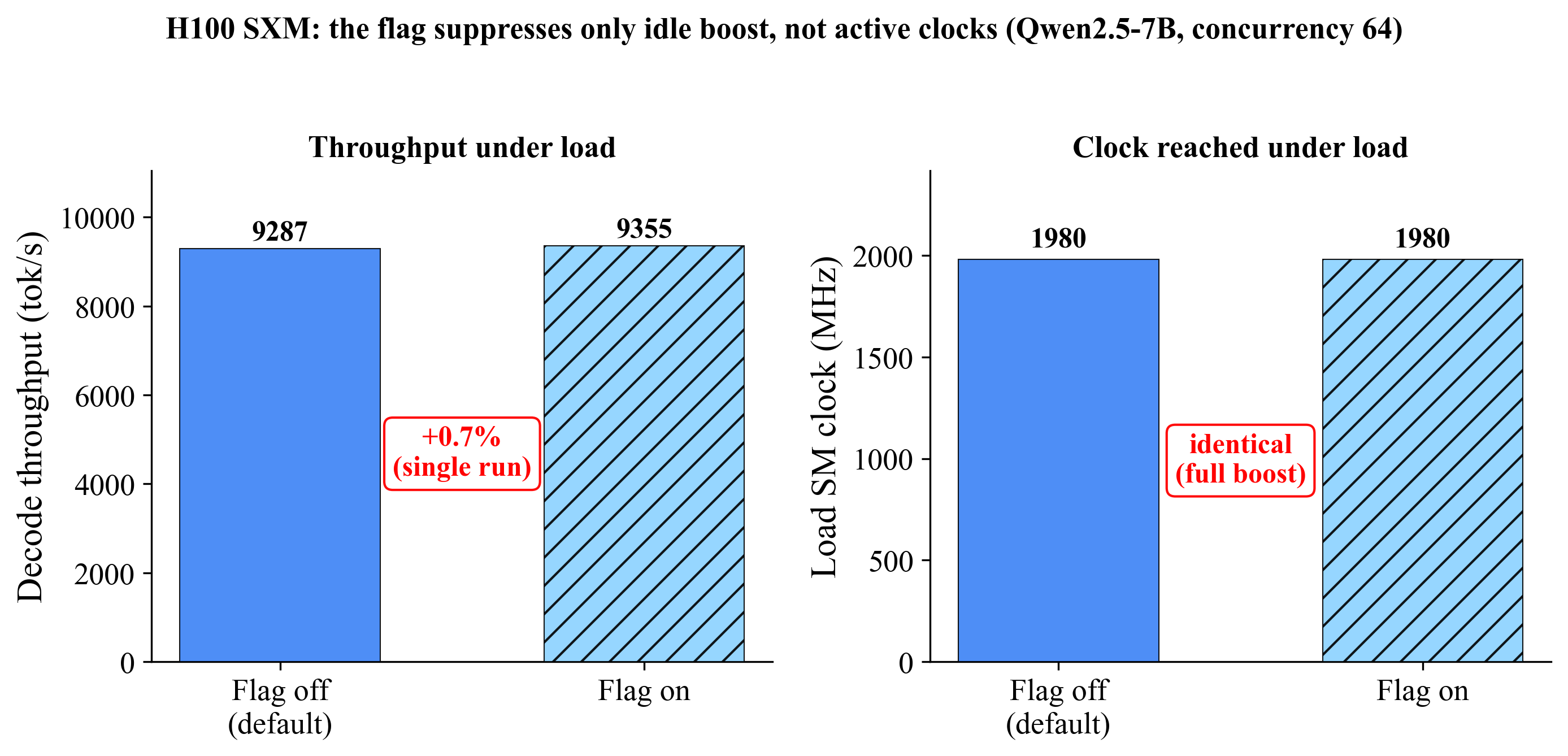}
    \caption{H100 decode throughput and load SM clock, flag off vs.\ on (Qwen2.5-7B, 512 requests at concurrency~64, single run). Throughput is 9287\,tok/s (baseline) and 9355\,tok/s (flag); both hold the 1980\,MHz load clock. The flag does not cap active clocks.}
    \label{fig:throughput}
\end{figure*}

\paragraph{Throughput under load.} To check that the flag does not cap active clocks, we ran a high-concurrency decode benchmark on the H100 (Qwen2.5-7B, 512 requests at concurrency~64; Figure~\ref{fig:throughput}). In a single run, aggregate decode throughput was 9287\,tok/s at baseline and 9355\,tok/s with the flag (a 0.7\% difference); repeated trials would be needed to characterize run-to-run variance. In both conditions the GPU held the same 1980\,MHz load SM clock---its full boost frequency---so the flag does not cap active clocks. It suppresses only the idle boost.

\subsection{Phase~4: Multi-GPU Scaling}
\label{sec:multi-gpu}

We extend the single-GPU findings to multi-GPU deployments, testing whether the parking tax compounds under tensor parallelism, data parallelism, and mixed configurations. We run a within-subject experiment on a single 4$\times$H100 80GB SXM node connected via NVLink (driver 580.126.09), testing 9~conditions (Table~\ref{tab:multi-gpu-conditions}) with Qwen2.5-7B (fp16) served by vLLM, following the same measurement protocol as Phase~2.

\begin{table}[tb]
\caption{Multi-GPU experimental conditions. All conditions use 4$\times$H100~SXM with NVLink. Flag refers to \perfboost\texttt{=1}.}
\label{tab:multi-gpu-conditions}
\small
\begin{tabular}{llrl}
\toprule
\textbf{Condition} & \textbf{Parallelism} & \textbf{GPUs} & \textbf{Flag} \\
\midrule
Bare idle & None & 4 & --- \\
TP=2 baseline & Tensor & 2 & Off \\
TP=2 flag on & Tensor & 2 & On \\
TP=4 baseline & Tensor & 4 & Off \\
TP=4 flag on & Tensor & 4 & On \\
DP=2 baseline & Data & 2 & Off \\
DP=2 flag on & Data & 2 & On \\
TP=2$\times$DP=2 baseline & Mixed & 4 & Off \\
TP=2$\times$DP=2 flag on & Mixed & 4 & On \\
\bottomrule
\end{tabular}
\end{table}

\paragraph{Setup.} Tensor parallelism (TP) shards the model across GPUs with NCCL all-reduce for inter-layer communication; data parallelism (DP) runs independent model replicas, each with its own CUDA context and no inter-GPU communication; the mixed TP=2$\times$DP=2 configuration runs two TP=2 groups as independent DP replicas (GPUs~0,1 and GPUs~2,3). Per condition we measured warm latency ($n=50$), a 5-minute DVFS decay curve at 1\,s intervals, steady-state idle power ($n=40$, 20-min phase), and the cold-start ramp (60\,s idle soak, then 1~cold request $+$ 4~recovery). Inter-condition cooldowns of 120\,s ensured thermal equilibrium.

\begin{figure}[t]
    \centering
    \includegraphics[width=\columnwidth]{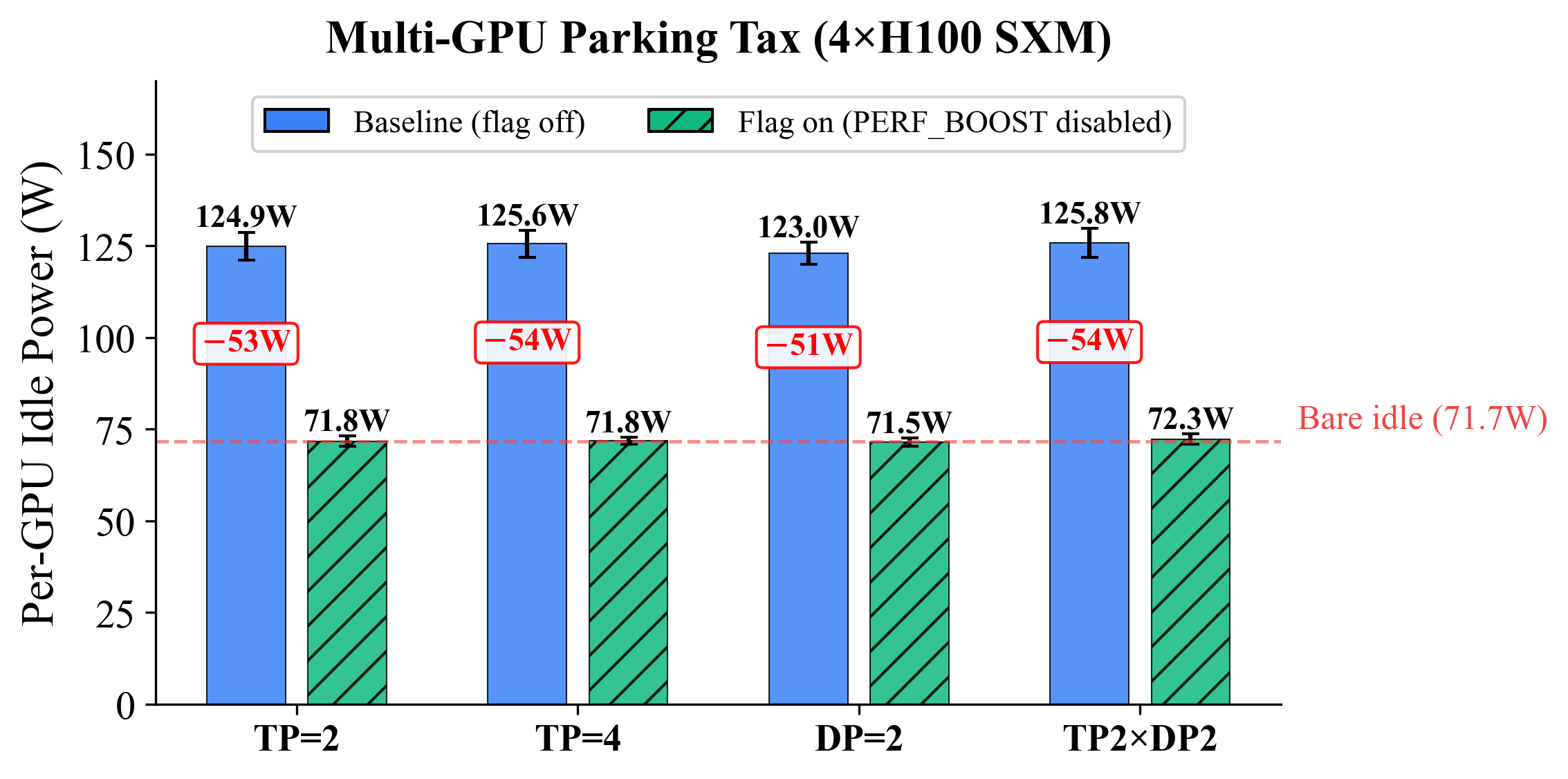}
    \caption{Per-GPU idle power across parallelism strategies on a 4$\times$H100~SXM node. Baseline (blue): ${\sim}$125\,W/GPU ($+$53\,W parking tax). Flag on (green, hatched): ${\sim}$72\,W/GPU (bare idle). The tax is consistent (${\sim}$52\,W/GPU) across TP, DP, and mixed configurations---the communication pattern does not matter.}
    \label{fig:multi-gpu-bars}
\end{figure}

\paragraph{Parking tax is per-context and linear with $N$.} Table~\ref{tab:multi-gpu-summary} presents the central result. Across all four parallelism strategies, the per-GPU parking tax is ${\sim}52$\,W---matching the single-GPU H100 measurement of 49.9\,W (Table~\ref{tab:cross}) within inter-device variance. The communication pattern does not matter: NCCL all-reduce (TP), independent contexts (DP), and mixed (TP$+$DP) all produce the same per-GPU overhead (Figure~\ref{fig:multi-gpu-bars}). A 4-GPU node with TP=4 wastes 215.6\,W idle; a TP=2$\times$DP=2 deployment wastes 216.4\,W. The tax scales as:
\begin{equation}
    P_{\text{park}}(N) = N \cdot \Delta P_{\text{DVFS}} \approx N \times 52\,\text{W (H100)}
    \label{eq:linear-scaling}
\end{equation}
where the ${\sim}52$\,W coefficient is specific to H100~SXM; on other architectures, substitute 26.3\,W (A100) or 67.2\,W (L40S) per Table~\ref{tab:cross}.

\begin{table}[tb]
\caption{Multi-GPU parking tax summary (4$\times$H100~SXM, Qwen2.5-7B fp16, driver 580.126.09). Per-GPU idle power is the mean across all GPUs in each condition ($n=40$ per GPU, 20-min phases). Tax is relative to bare idle (71.7\,W/GPU). Warm latency: $n=50$ back-to-back requests, 128-token prompt/completion. Cold req\,1: first request after 60\,s idle soak (32-token completion). Penalty: cold req\,1 $-$ recovery mean.}
\label{tab:multi-gpu-summary}
\small
\begin{tabular}{@{}lrrrrr@{}}
\toprule
\textbf{Condition} & \makebox[3.5em][r]{\textbf{Per-GPU}} & \makebox[3em][r]{\textbf{Tax/}} & \makebox[3.5em][r]{\textbf{Warm}} & \makebox[4em][r]{\textbf{Cold}} & \makebox[4em][r]{\textbf{Penalty}} \\
 & \makebox[3.5em][r]{\textbf{(W)}} & \makebox[3em][r]{\textbf{GPU}} & \makebox[3.5em][r]{\textbf{(ms)}} & \makebox[4em][r]{\textbf{req\,1 (ms)}} & \makebox[4em][r]{\textbf{(ms)}} \\
\midrule
Bare idle & 71.7 & --- & --- & --- & --- \\
\midrule
TP=2 baseline & 124.9 & $+$53.2 & 340.3 & 133.0 & $+$5.4 \\
TP=2 flag on & 71.8 & $+$0.0 & 352.9 & 302.9 & $+$175.5 \\
\midrule
TP=4 baseline & 125.6 & $+$53.9 & 237.7 & 103.5 & $+$13.8 \\
TP=4 flag on & 71.8 & $+$0.1 & 250.6 & 258.5 & $+$156.5 \\
\midrule
DP=2 baseline & 123.0 & $+$51.3 & 531.4 & 198.0 & $+$5.0 \\
DP=2 flag on & 71.5 & $-$0.2 & 520.7 & 344.1 & $+$153.4 \\
\midrule
TP2$\times$DP2 base & 125.8 & $+$54.1 & 352.9 & 134.3 & $+$7.2 \\
TP2$\times$DP2 flag & 72.3 & $+$0.6 & 353.1 & 272.4 & $+$144.9 \\
\bottomrule
\end{tabular}
\end{table}

\paragraph{Flag eliminates the tax under NCCL.} Every flag-on condition in Table~\ref{tab:multi-gpu-summary} drops per-GPU idle power to within 0.6\,W of bare idle (71.7\,W). The CUDA driver respects the flag per-GPU regardless of active NCCL communication groups, NVLink topology, or the number of CUDA contexts on the node. Table~\ref{tab:multi-gpu-per-gpu} shows the per-GPU power breakdown; GPU-to-GPU variance within each condition is small ($<$10\,W baseline, $<$4\,W flag on), confirming the result is not driven by outlier GPUs.

\begin{table}[tb]
\caption{Per-GPU idle power detail (4$\times$H100~SXM, $n=40$ per GPU). Baseline conditions show the characteristic GPU-to-GPU spread; flag-on conditions collapse all GPUs to bare idle. Entries marked --- indicate GPUs not used in that condition.}
\label{tab:multi-gpu-per-gpu}
\small
\begin{tabular}{@{}lrrrr@{}}
\toprule
\textbf{Condition} & \textbf{GPU\,0 (W)} & \textbf{GPU\,1 (W)} & \textbf{GPU\,2 (W)} & \textbf{GPU\,3 (W)} \\
\midrule
TP=2 baseline & 128.7 & 121.1 & --- & --- \\
TP=2 flag on & 73.1 & 70.5 & --- & --- \\
TP=4 baseline & 127.8 & 121.1 & 130.3 & 123.1 \\
TP=4 flag on & 72.7 & 70.5 & 72.6 & 71.4 \\
DP=2 baseline & 126.0 & 120.0 & --- & --- \\
DP=2 flag on & 72.5 & 70.4 & --- & --- \\
TP2$\times$DP2 base & 129.9 & 120.8 & 129.3 & 123.3 \\
TP2$\times$DP2 flag & 74.3 & 70.6 & 72.9 & 71.6 \\
\bottomrule
\end{tabular}
\end{table}

\paragraph{Warm latency unaffected.} Steady-state inference latency with the flag enabled is within noise of baseline across all configurations: TP=2 increases by 3.7\% (340$\to$353\,ms), TP=4 by 5.4\% (238$\to$251\,ms), and DP=2 decreases slightly (531$\to$521\,ms). Warm latency is similar to baseline in this retest (one run per condition), and these differences do not indicate a systematic penalty from the flag during active inference.

\begin{figure}[t]
    \centering
    \includegraphics[width=\columnwidth]{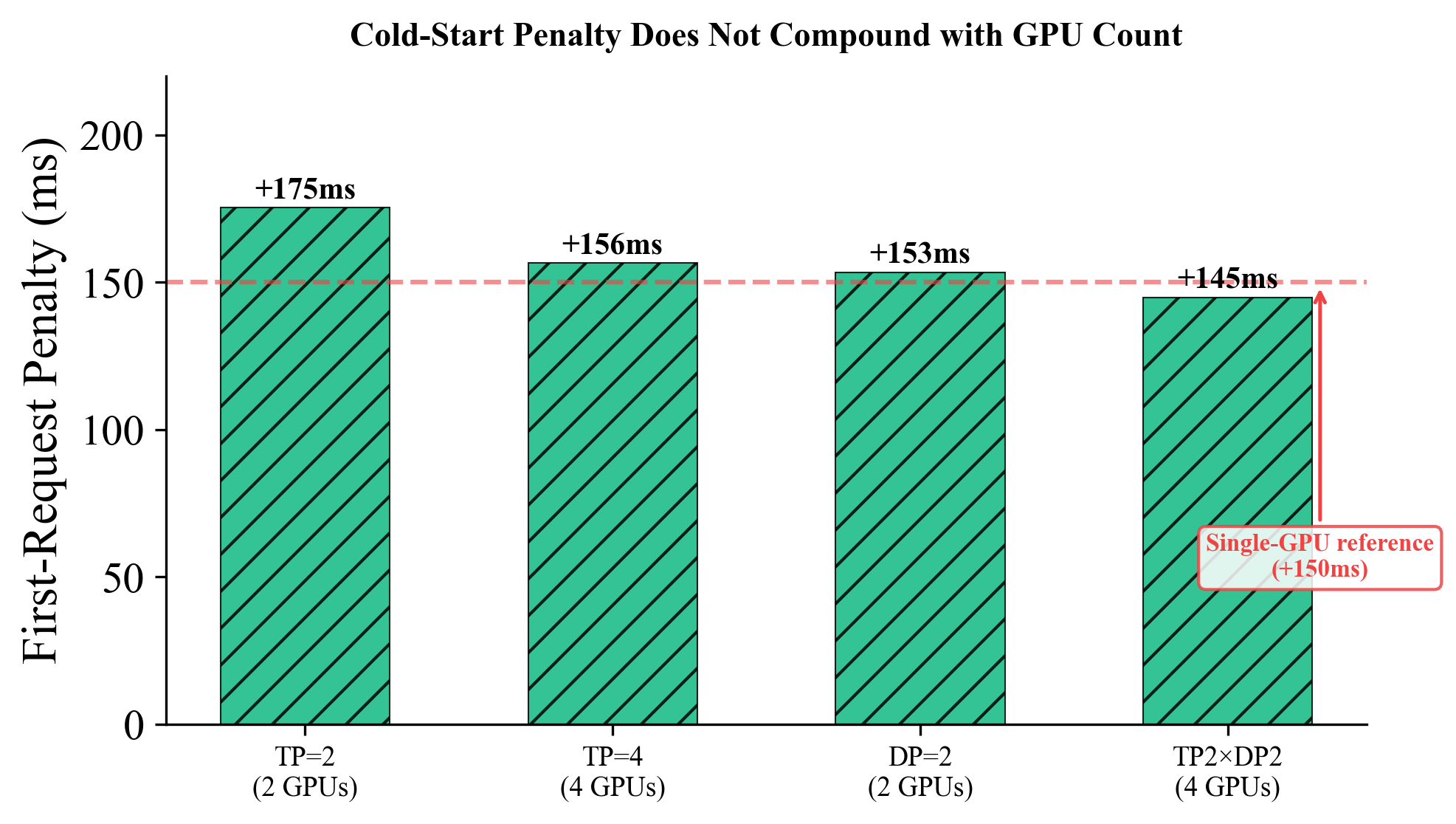}
    \caption{First-request DVFS ramp penalty with the flag enabled across multi-GPU configurations. All penalties fall within the range of the single-GPU baseline ($+$150\,ms). The penalty does not compound with GPU count: 4-GPU configurations (TP=4, TP2$\times$DP2) show penalties comparable to 2-GPU configurations (TP=2, DP=2).}
    \label{fig:cold-start-penalty}
\end{figure}

\paragraph{Cold-start penalty does not compound with $N$.} The first-request DVFS ramp penalty with the flag enabled is $+$145--176\,ms across all multi-GPU configurations (Table~\ref{tab:multi-gpu-summary}), matching the single-GPU penalty of ${\sim}$150\,ms (\S\ref{sec:perf-boost}, Figure~\ref{fig:cold-start-penalty}). Under tensor parallelism, the first CUDA kernel launch triggers NCCL collective operations that act as implicit barriers, causing all GPUs to begin their DVFS ramp simultaneously. Since all $N$ GPUs ramp in parallel, the user-facing latency is bounded by the slowest single GPU's ramp time, not $N$ sequential ramps. Under data parallelism, each replica ramps independently on its first request---there is no cross-replica synchronization cost. In both cases, the penalty is constant with GPU count.

\begin{figure}[t]
    \centering
    \includegraphics[width=\columnwidth]{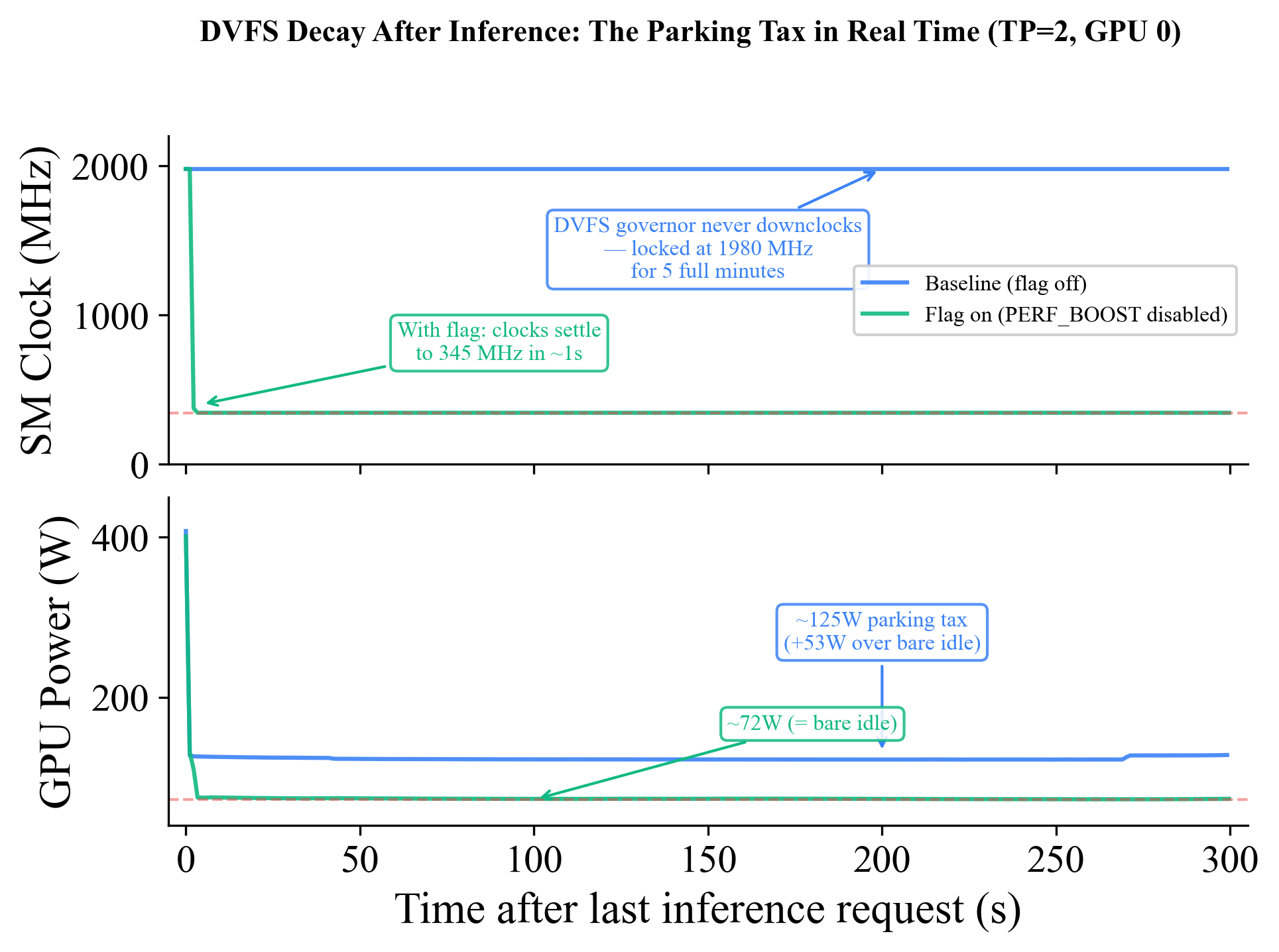}
    \caption{DVFS decay after last inference request (TP=2, GPU~0). Top: SM clock frequency. Bottom: GPU power. The trace begins at active-inference power (${\sim}$400\,W under load). Without the flag, clocks remain locked at 1980\,MHz for 5~full minutes and power settles at the CUDA-active idle plateau (${\sim}$125\,W)---the parking tax persists indefinitely. With the flag, clocks drop to bare idle (345\,MHz) within 2\,s, and power transitions from active (${\sim}$400\,W) through the CUDA-active plateau (${\sim}$125\,W) to bare idle (${\sim}$72\,W) in $<$4\,s.}
    \label{fig:decay-curve}
\end{figure}

\paragraph{DVFS decay curve.} Without the flag, SM clocks remain locked at 1980\,MHz for the entire 5-minute recording window after the last request---the DVFS governor never self-corrects (Figure~\ref{fig:decay-curve}, top). With the flag enabled, clocks drop from 1980$\to$345\,MHz within 2\,s of the last request, and power transitions from the CUDA-active idle plateau (${\sim}$125\,W) to bare idle (${\sim}$72\,W) in $<$4\,s (Figure~\ref{fig:decay-curve}, bottom). The flag is set once at process start and requires no runtime toggling---the DVFS governor handles idle-to-active transitions automatically when CUDA kernels are launched.

\section{Industry-Scale Impact}

Let $N$ denote total datacenter GPUs, $\rho$ the average utilization, and $T_{\text{year}} = 8{,}760$\,h:
\begin{equation}
    E_{\text{park}} = N(1 - \rho) \cdot \bar{P}_{\text{park}} \cdot T_{\text{year}}
    \label{eq:industry}
\end{equation}
NVIDIA shipped $\sim$3.76M datacenter GPUs in 2023~\cite{nvidia-shipments}. Taking indicative assumptions of $\rho = 0.65$ (35\% idle~\cite{flexera}) and $\bar{P}_{\text{park}} = 40$\,W (fleet-weighted average across architectures), the base estimate is $E_{\text{park}} = 462$\,GWh/year.

\paragraph{Sensitivity analysis.} We vary each parameter across its plausible range (Table~\ref{tab:sensitivity}) to bound the estimate. The dominant uncertainty is fleet size and utilization rate, not the parking tax itself (which we measure directly).

\begin{table}[tb]
\caption{Sensitivity of industry impact to key assumptions.}
\label{tab:sensitivity}
\small
\begin{tabular}{lrrr}
\toprule
\textbf{Parameter} & \textbf{Low} & \textbf{Base} & \textbf{High} \\
\midrule
Fleet size & 2.0M & 3.76M & 6.0M \\
Utilization $\rho$ & 50\% & 65\% & 80\% \\
$\bar{P}_{\text{park}}$ (W) & 26.3 (A100) & 40 & 67.2 (L40S) \\
\midrule
$E_{\text{park}}$ (GWh/yr) & 92 & 462 & 1{,}766 \\
\bottomrule
\end{tabular}
\end{table}

The plausible range spans 92--1{,}766\,GWh/year, with the base case at 462\,GWh/year ($\sim$180\,kT CO$_2$ at US grid average). Even the conservative lower bound (92\,GWh) exceeds the annual electricity consumption of a small city. The range reflects uncertainty in fleet composition and duty cycles---not in the parking tax measurement itself, which is tightly bounded by our experiments.

\section{Scheduler Evaluation}
\label{sec:scheduler}

\emph{Note:} The \perfboost{} flag (\S\ref{sec:perf-boost}) is the preferred mitigation when available (NVIDIA GPUs, driver in the 580 series; tested at $\geq$580.105.08). The breakeven scheduler below is a \textbf{fallback} for deployments without flag access---older drivers, non-NVIDIA hardware, or environments where the flag cannot be set.

To demonstrate the practical utility of the breakeven model, we implement a proof-of-concept scheduler that uses $T^*$ (Equation~\ref{eq:breakeven}) to make keep-warm/evict decisions. We compare three policies over 24-hour simulations:

\begin{enumerate}
    \item \textbf{Always-On}: Model stays loaded (industry default).
    \item \textbf{Fixed time-to-live (TTL)}: Evict after a fixed idle timeout (5, 15, or 30 minutes).
    \item \textbf{Breakeven}: Evict after $T^* = P_{\text{load}} \cdot t_{\text{load}} / P_{\text{park}}$ seconds of idle time.
\end{enumerate}

We evaluate on three synthetic traffic patterns: steady Poisson (5\,req/hr), bursty (alternating 2 and 60\,req/hr), and diurnal (sinusoidal, peak 30\,req/hr). On H100 with standard PyTorch loading, $T^* = 271$\,s (4.5~min).

\paragraph{Results.} The breakeven policy saves 8.2--23.0\% energy versus always-on across traffic patterns (Table~\ref{tab:scheduler}). On bursty traffic---the most realistic pattern for production LLM serving---the breakeven policy saves 23.0\% with only 48 cold starts over 24 hours (4.5\,s average added latency per request). The breakeven policy matches or outperforms all fixed TTLs because $T^*$ is derived from measured hardware parameters rather than guessed.

\begin{table}[tb]
\caption{Scheduler simulation results (H100, 24h, standard PyTorch loader). Energy savings vs.\ always-on baseline (2{,}921\,Wh).}
\label{tab:scheduler}
\small
\begin{tabular}{lrrr}
\toprule
\textbf{Policy} & \textbf{Energy (Wh)} & \textbf{Savings} & \textbf{Cold starts} \\
\midrule
\multicolumn{4}{l}{\emph{Low traffic (5 req/hr, Poisson)}} \\
Always-On & 2{,}921 & --- & 1 \\
TTL 5\,min & 2{,}407 & 17.6\% & 78 \\
Breakeven ($T^*$=4.5\,min) & 2{,}392 & 18.1\% & 81 \\
\midrule
\multicolumn{4}{l}{\emph{Bursty traffic (2/60 req/hr)}} \\
Always-On & 2{,}921 & --- & 1 \\
TTL 5\,min & 2{,}264 & 22.5\% & 47 \\
Breakeven ($T^*$=4.5\,min) & 2{,}248 & 23.0\% & 48 \\
\midrule
\multicolumn{4}{l}{\emph{Diurnal (peak 30 req/hr)}} \\
Always-On & 2{,}921 & --- & 1 \\
TTL 5\,min & 2{,}671 & 8.6\% & 87 \\
Breakeven ($T^*$=4.5\,min) & 2{,}682 & 8.2\% & 100 \\
\bottomrule
\end{tabular}
\end{table}

\paragraph{Cross-architecture behavior.} The breakeven time varies by architecture: $T^* = 271$\,s (H100), 513\,s (A100), and 203\,s (L40S). The L40S, with the highest parking tax (67.2\,W), has the shortest breakeven time and the most to gain from eviction. With fast model loading (ServerlessLLM, $t_{\text{load}} = 8$\,s), $T^*$ drops to 48\,s (H100), enabling aggressive eviction even at moderate request rates.

\section{Discussion}

\paragraph{Cross-architecture consistency.} The parking tax is a step function of CUDA context on H100 (Hopper), A100 (Ampere), and L40S (Ada Lovelace), and on Blackwell (B200)---four microarchitectures with different memory technologies, process nodes, and TDP budgets. The magnitude varies (26--67\,W), but the structure is identical: $\partial P / \partial C \gg \partial P / \partial V$. With $n = 1$ device per architecture in Phase~2 (with an $n=18$ protocol on the B200), we cannot claim universal invariance, but the consistency---combined with the physics argument of Equation~\ref{eq:mem-prediction}---points to a design-space constraint rather than a device-specific artifact. Phase~1 bounds inter-device variation directly: regression across all 14~H100 GPUs yields slopes indistinguishable from zero on every device, even as intercepts vary by $\sim$23\,W from silicon binning and cooling. This pattern is consistent with $\beta \approx 0$ being device-invariant rather than an artifact of any particular GPU.

\paragraph{Relationship to execution-idle.} Lei et al.~\cite{lei-idle} find that execution-idle accounts for 19.7\% of in-execution time and 10.7\% of energy across 756~GPUs. Execution-idle is a superset of the zero-activity state we measure---it includes I/O stalls, synchronization barriers, and data loading pauses. Our parking tax is the dominant sub-component when GPU activity is strictly zero. On the L40S, our CUDA context overhead (67.2\,W) accounts for the bulk of the gap between their deep-idle ($\sim$35\,W) and execution-idle ($\sim$110\,W) power levels. Their finding that serving workloads spend 61\% of in-execution time idle, combined with our result that idle power is model-size-independent, points to small always-on deployments as the largest per-GPU energy waste.

\paragraph{Why $\Delta P_{\text{DVFS}}$ varies.} The L40S has the highest context overhead (67.2\,W, 19\% of TDP) despite half the TDP of the H100 (350\,W vs.\ 700\,W), because it boosts to 2520\,MHz (vs.\ 1980\,MHz for H100, 1410\,MHz for A100). Higher idle clock frequency means higher idle power.

\paragraph{Why boost is the default, and when it is the wrong policy.}
\label{sec:default-regime}
Boost-on-context is consistent with a default that favors first-request latency, not a defect. Pre-ramping clocks the instant a context is created eliminates the DVFS ramp on the first kernel, which benefits \emph{sporadic interactive} workloads where any request may follow a long idle gap and tail latency is contractual. The trade is continuous idle power for a one-time latency saving, and it can be quantified. Keeping boost active through an idle gap of duration $t_{\text{gap}}$ costs
\begin{equation}
    E_{\text{default}} = \Delta P_{\text{DVFS}} \cdot t_{\text{gap}}
    \label{eq:default-cost}
\end{equation}
to save a single ramp of $r \approx 150$\,ms (Table~\ref{tab:cold-ramp}) on at most one request. On an H100 ($\Delta P_{\text{DVFS}} \approx 50$\,W), a 1\,s gap spends ${\sim}50$\,J to shave 150\,ms off one request; a 60\,s gap spends ${\sim}0.83$\,Wh. The default pays energy continuously to hedge a bounded, one-time latency cost.

The decision rule follows. Let $g$ be the fraction of requests that arrive after an idle gap long enough for clocks to have dropped, and let the SLO tolerate added tail latency $\tau_{\text{SLO}}$. The default is justified only when the ramp would violate the SLO \emph{and} cold arrivals are common ($r > \tau_{\text{SLO}}$ and $g$ large). For always-on serving---the segment that, by Phase~1 and Lei et al.~\cite{lei-idle}, dominates GPU idle time---requests are batched and warm, $g$ is small, and a ${\sim}150$\,ms ramp that is gone by the second request rarely threatens a realistic SLO. In that regime the default is the wrong energy policy and the flag should be on. For sub-second-SLO interactive endpoints with mostly-cold arrivals, the vendor default remains defensible. A single global default cannot be correct for both regimes; the flag exists because the appropriate choice is per-deployment, and the parking tax is the cost of applying a latency-oriented default to always-on serving.

\paragraph{Prescription: flag at process start.}
\label{sec:prescription}
Based on our measurements, setting \perfboost\texttt{=1} in the process environment before launching the inference server appears to be the cheapest available mitigation. The flag modifies the DVFS governor at CUDA context creation. No runtime toggling is needed:
\begin{enumerate}
    \item \textbf{While serving}: full boost frequency, no latency penalty at any percentile (Table~\ref{tab:multi-gpu-summary}). CUDA graphs work correctly.
    \item \textbf{While idle}: clocks drop to bare idle (${\sim}72$\,W/GPU on H100) within 2\,s of the last kernel.
    \item \textbf{On next request}: one-time ${\sim}150$\,ms ramp, constant regardless of GPU count or parallelism strategy. Gone by request~2; a warmup probe eliminates it entirely.
\end{enumerate}
Savings scale linearly with $N$ (Equation~\ref{eq:linear-scaling}): 4-GPU node saves ${\sim}215$\,W, 8-GPU node saves ${\sim}430$\,W. At 30\% idle fraction~\cite{flexera,lei-idle}, a 1{,}000-GPU cluster saves ${\sim}16$\,kW (${\sim}140$\,MWh/year). Implementation: one environment variable in the container spec or launcher script, with an optional keep-alive probe for latency-sensitive deployments.

\paragraph{Pipeline parallelism and multi-node deployments.}
Our experiments cover TP and DP on a single NVLink node. Two extrapolations follow. Under pipeline parallelism (PP), stages communicate via point-to-point send/recv rather than all-reduce; the mechanism we identify is per-CUDA-context and independent of communication pattern (\S\ref{sec:multi-gpu}), so the per-GPU tax should be unchanged, though we have not measured it directly. For multi-node deployments, the DVFS governor on each GPU responds to local SM activity, not the interconnect fabric, so InfiniBand or RoCE should not change the per-GPU result. Under this hypothesis, a 64-GPU deployment across 8 nodes would waste $64 \times 52 \approx 3{,}328$\,W idle and recover the same via the flag. Both extrapolations are testable; we leave direct validation as future work.

\paragraph{Adoption barriers.}
The flag has been available since November 2025 but has not seen adoption. We cannot answer why empirically, but the structure of the inference-serving market suggests an explanation: the engineers configuring inference engines are typically not the engineers paying the power bill. Cloud GPU tenants pay flat hourly rates with no direct visibility into idle power; the cost falls on infrastructure operators---cloud providers running managed endpoints, inference companies operating GPU fleets, enterprises with on-premise clusters. If this principal-agent gap is real, native support for the flag in inference-serving frameworks (vLLM~\cite{vllm}, SGLang, TensorRT-LLM) could reduce this friction and let the operators with the incentive to act do so at fleet scale.

\paragraph{Design directions.} (1)~\textbf{CUDA context power management}: The \perfboost{} flag returns idle-with-context power to the bare-idle floor, recovering roughly 95--100\% of the overhead across all four architectures (H100, A100, L40S, and B200) (\S\ref{sec:perf-boost}, \S\ref{sec:multi-gpu}). No steady-state latency penalty is measured on the H100 and A100, and no throughput penalty in an H100 load test. The parking tax is a software policy artifact, not a hardware constraint. Orchestrators (Kubernetes device plugins, vLLM launcher scripts) could set the flag by default. (2)~\textbf{Context pooling}: MPS~\cite{nvidia-mps} to share CUDA contexts and amortize overhead; MIG~\cite{nvidia-mig} to isolate tenants while sharing the DVFS cost. (3)~\textbf{Size-aware scheduling}: Preferentially cold-start small models (fast reload, same parking cost). (4)~\textbf{Breakeven-aware eviction}: A $T^*$-based policy saves 8--23\% energy (\S\ref{sec:scheduler}) using only hardware parameters we measure. The breakeven policy slightly underperforms fixed TTLs on gradual traffic ramps (diurnal pattern, Table~\ref{tab:scheduler}) because threshold-based eviction oscillates near the crossover rate; adaptive smoothing or hysteresis could address this.

\paragraph{Limitations.} We test one device per architecture in Phase~2 under a controlled within-subject design that isolates VRAM allocation as the sole manipulated variable, eliminating between-device confounders (silicon binning, cooling, power supply). Phase~1 complements this with \num{335267} samples across 14~H100 devices, bounding inter-device slope variation to below measurement sensitivity. The three primary architectures (H100, A100, L40S) use $n=40$ per phase (20-minute recording windows), while the Blackwell B200 uses a shorter $n=18$ sweep (3-minute phases). The primary Phase~2 experiments use \texttt{torch.empty} tensors; we validate with a real Qwen2.5-7B model, which on the H100 idles within 0.03\,W of an equal-context empty allocation, with the L40S real-model idle corroborating the context-active level (\S4.3, Table~\ref{tab:validation}). Cold-start power traces are captured on the H100; load times are measured on all three. Larger models may exhibit different loading profiles, though the VRAM-independence result is tested up to 72\,GB with controlled allocations. While \texttt{nvidia-smi} samples power intermittently~\cite{burtscher-power}, the low within-phase variance ($<$1.5\,W) on all devices confirms that idle power is stable enough for unbiased mean estimation. We do not evaluate MIG~\cite{nvidia-mig}, MPS~\cite{nvidia-mps}, or power capping (\texttt{nvidia-smi -pl}) configurations; these may alter the DVFS behavior we observe. Our \perfboost{} latency and throughput evaluation covers the H100 and A100 (with cross-variant validation on PCIe vs.\ SXM4, \S\ref{sec:perf-boost}); on the L40S and B200 we measured only idle-power recovery (the B200 with an $n=18$ protocol). Behavior under multi-tenant workloads may differ. Steady-state latency measurements ($n=50$) on the H100 and A100 show no penalty at any percentile, and a single-run H100 throughput test shows none under load, though repeated trials would be needed to characterize variance. A focused retest with a 60\,s idle soak confirms a one-time first-request DVFS ramp ($+$147--220\,ms) on both architectures, gone by request~2; whether this ramp varies with idle duration or model size is untested. Our multi-GPU experiments (\S\ref{sec:multi-gpu}) test tensor and data parallelism on a single 4$\times$H100~SXM node with NVLink; whether the linear scaling holds under pipeline parallelism, on 8-GPU nodes, across NVSwitch fabrics, in multi-node deployments over InfiniBand/RoCE, or on non-Hopper architectures under multi-GPU configurations is untested. We measure Qwen2.5-7B (7B parameters); larger models requiring more GPUs may exhibit different NCCL initialization or cold-start behavior. The flag-at-start prescription assumes the flag can be set per-process without side effects on co-located workloads; interaction with job schedulers, container runtimes, or MIG partitions is unexplored. Because \perfboost{} is an environment variable read at CUDA context creation, it cannot be toggled on a running process---deployments requiring mid-process behavior changes would need to restart the inference server or recreate the CUDA context. The scheduler evaluation uses synthetic traces; validation on production LLM serving logs---such as the cluster-scale telemetry collected by Lei et al.~\cite{lei-idle}---is a natural next step.

\section{Conclusion}

The model parking tax (26--67\,W per GPU, depending on architecture) is a fixed cost of the CUDA context, not a variable cost of model size. Across four architectures spanning HBM3, HBM2e, GDDR6, and HBM3e (Blackwell), in production and in controlled experiment, all four VRAM slopes are below 0.02\,W/GB, bounded well below the 0.1\,W/GB relevance threshold by two one-sided tests. NVIDIA's \perfboost{} flag returns idle-with-context power to the bare-idle floor, recovering roughly 95--100\% of the overhead, with no steady-state latency penalty measured on the H100 and A100; the only cost is a transient ${\sim}150$\,ms ramp on the first request after idle, gone by the second request. The tax scales linearly with GPU count across all parallelism strategies tested, and the cold-start penalty does not compound with $N$---NCCL barriers synchronize all GPUs to ramp in parallel. With the flag set at process start, clocks drop to bare idle within 2\,s of the last request automatically, requiring no runtime intervention. At industry scale, a mid-range scenario places the recoverable overhead near \num{460}\,GWh/year (${\sim}180$\,kT CO$_2$), within a wider 92--\num{1766}\,GWh/year range that reflects fleet and duty-cycle uncertainty.

Setting the flag at process start lets the DVFS governor handle idle-active transitions automatically. The audience for this prescription is infrastructure operators who bear the energy cost, not application developers who pay flat hourly rates. Native integration into inference-serving frameworks would help close this gap. Whether the prescription holds at scale, under multi-tenant workloads, and across architectures we did not test remains the open question.

\paragraph{Future work.} The breakeven model (Equation~\ref{eq:breakeven}) assumes known arrival rates. Replacing the static threshold with a predictive controller trained on production traces---enabled by the architecture-independent structure of $P_{\text{park}}$---is a direct next step. Validation on cluster-scale telemetry such as Lei et al.'s~\cite{lei-idle} would ground these estimates in real demand uncertainty.

\paragraph{Reproducibility.} All data, analysis scripts, and experiment code are available at \href{https://github.com/8bitai/gpu-parking-tax}{https://github.com/8bitai/gpu-parking-tax}.

\begin{acks}
Generative AI tools (Anthropic's Claude, used through Claude Code) assisted with editing the prose, writing the figure-generation and analysis scripts in the companion repository, and auditing numerical consistency between the text, tables, and figures. The experimental design, all measurements, the statistical analysis, and the interpretation of the results are the author's, who takes full responsibility for the content.
\end{acks}

\bibliographystyle{ACM-Reference-Format}
\def\bibfont{\normalsize}

\end{document}